\documentclass{naturep}

\usepackage{amssymb}
\usepackage{amsmath}
\usepackage{graphicx}
\usepackage{algorithmicx}
\usepackage{algorithm}
\usepackage[noend]{algpseudocode}
\usepackage{bbm}
\usepackage{lineno}
\usepackage[margin=0.6in]{geometry}
\usepackage{ragged2e}
\usepackage{setspace}
\usepackage{hyperref}
\usepackage{caption}
\usepackage{anyfontsize}
\usepackage{setspace}
\usepackage{float}

\makeatletter
\let\saved@includegraphics\includegraphics
\AtBeginDocument{\let\includegraphics\saved@includegraphics}

\makeatother

\title{\begin{flushleft}{\begin{spacing}{1}Data Efficient and Weakly Supervised Computational Pathology on Whole Slide Images\end{spacing}}\end{flushleft}}

\begin{document}

\maketitle
\begin{spacing}{1.8}
\vspace{-15mm}
\noindent Ming Y. Lu$^{1,2}$, Drew F. K. Williamson$^{1,5}$, Tiffany Y. Chen$^{1,5}$, Richard J. Chen$^{1,4}$, Matteo Barbieri$^{1,2}$,\\ and Faisal Mahmood$^{*1,2,3}$
\begin{affiliations}
 \item Department of Pathology, Brigham and Women's Hospital, Harvard Medical School, Boston, MA
 \item Cancer Program, Broad Institute of Harvard and MIT, Cambridge, MA 
 \item Cancer Data Science, Dana-Farber Cancer Institute, Boston, MA
 \item Department of Biomedical Informatics, Harvard Medical School, Boston, MA
 \item Contributed Equally
 \end{affiliations}
 
 \noindent\textbf{Code / Package:} http://github.com/mahmoodlab/CLAM\\
 \textbf{Interactive Demo:} http://clam.mahmoodlab.org\\
 
  \end{spacing}
\begin{spacing}{1.4}
\noindent\textbf{*Correspondence:}\\ 
Faisal Mahmood \\
60 Fenwood Road, Hale Building for Transformative Medicine\\
Brigham and Women's Hospital, Harvard Medical School\\
Boston, MA 02445\\
faisalmahmood@bwh.harvard.edu
\end{spacing}

\newpage
\noindent\textbf{\large{Abstract}}
\vspace{-5mm}
\begin{spacing}{1.5}
\noindent The rapidly emerging field of computational pathology has the potential to enable objective diagnosis, therapeutic response prediction and identification of new morphological features of clinical relevance. However, deep learning-based computational pathology approaches either require manual annotation of gigapixel whole slide images (WSIs) in fully-supervised settings or thousands of WSIs with slide-level labels in a weakly-supervised setting. Moreover, whole slide level computational pathology methods also suffer from domain adaptation, interpretability and visualization issues. These challenges have prevented the broad adaptation of computational pathology for clinical and research purposes. Here we present CLAM - Clustering-constrained Attention Multiple instance learning (\url{https://github.com/mahmoodlab/CLAM}), an easy-to-use, high-throughput and interpretable WSI-level processing and learning method that only requires slide-level labels while being data efficient, adaptable and capable of handling multi-class subtyping problems. CLAM is a deep-learning based weakly-supervised method that uses attention-based learning to automatically identify sub-regions of high diagnostic value in order to accurately classify the whole slide, while also utilizing instance-level clustering over the representative regions identified to constrain and refine the feature space. In three separate analyses, we demonstrate the data efficiency and adaptability of CLAM and its superior performance over standard weakly-supervised classification. We demonstrate that CLAM models are interpretable and can be used to identify well-known and new morphological features without using any spatial labels during training. We further show that models trained using CLAM are adaptable to independent test cohorts, cell phone microscopy images, and varying slide tissue content. CLAM is a flexible, general purpose, and adaptable method that can be used for a variety of different computational pathology tasks in both clinical and research settings. 
\end{spacing}

\newpage

\begin{spacing}{1.42}


Advances in digital pathology and artificial intelligence have presented the potential to analyze gigapixel whole slide images (WSIs) for objective diagnosis, prognosis, and therapeutic response prediction\cite{bera2019artificial,niazi2019digital}. Besides the immediate clinical benefits\cite{hollon2020near,kather2019deep,bulten2020automated,strom2020artificial}, computational pathology has demonstrated promise in a variety of different tasks by quantifying the tissue microenviornment\cite{schapiro2017histocat,moen2019deep,mahmood2019deep,graham2019hover,saltz2018spatial,javed2020cellular}, conducting integrative image-omic analysis\cite{mobadersany2018predicting,heindl2018microenvironmental,yuan2012quantitative,lazar2017comprehensive}, identifying morphological features of prognostic relevance\cite{beck2011systematic, yamamoto2019automated}, and associating morphologies with response and resistance to treatment\cite{pell2019use}. 

\vspace{-4mm}
Though deep learning\cite{lecun2015deep,esteva2019guide} has revolutionized medical imaging by solving many image classification and prediction tasks \cite{esteva2017dermatologist,poplin2018prediction,mckinney2020international,mitani2020detection,shen2019patient}, whole-slide imaging is a complex domain with several unique challenges. Deep learning-based computational pathology approaches require either manual annotation of gigapixel whole slide images (WSIs) in fully-supervised settings or large datasets with slide-level labels in a weakly-supervised setting. Since slide-level labels may only correspond to a tiny region of the large gigapixel image, to saliently localize these “needles in a haystack,” most approaches have relied on pixel, patch, or region of interest-level (ROI-level) annotations made by pathologists\cite{bejnordi2017diagnostic,chen2019augmented,nagpal2019development}. While promising results have been reported by assigning the same label to every patch in a WSI\cite{natmedlung}, this approach suffers from noisy training labels and is not applicable to problems that may have limited tumor content (\textit{e.g.,} micro-metastasis). Recent work has demonstrated exceptional clinical grade performance using slide-level labels for training binary classifiers for patient stratification in a weakly-supervised setting\cite{campanella2019clinical}. This methodology, however, required thousands of WSIs to achieve performance comparable with patch-based classifiers. Such large datasets are difficult to curate for rare diagnoses where only a handful of examples may exist or for clinical trials where it may be useful to predict outcome from a small cohort of patients. Moreover, most weakly-supervised whole slide classification methods are not directly applicable to multiclass tissue subtyping problems where normal tissue slides may not be available. Additionally, the performance of deep learning diagnostic models, especially those trained using patch-level supervision, have been shown to suffer when tested on data from different sources and imaging devices \cite{campanella2019clinical, natmedlung}. Such methods also need to be interpretable, with the capability to saliently localise regions used to make predictive determinations. In summary, for the broader adaptation of computational pathology in both clinical and research settings there is a need for methods that do not require manual ROI extraction, or pixel/patch-level labeling while still being data efficient, interpretable, adaptable, and capable of handling multiclass subtyping problems. 
\vspace{-4mm}
Here we propose CLAM as a novel, high-throughput deep learning framework that addresses the five key challenges with whole slide-level computational pathology outlined above. In three separate analyses (renal cell carcinoma subtyping, non-small cell lung cancer subtyping, and lymph node metastasis detection) using both publicly available datasets as well as independent test cohorts, we show that our approach is data efficient and can achieve high performance across different tasks while using systematically decreasing training labels. We demonstrate the adaptability of CLAM by showing that models trained on tissue resection WSIs can directly be applied to biopsy WSIs as well as microscopy images taken with a consumer grade cellphone using data from independent test cohorts. We also demonstrate that CLAM can generalize to multi-class classification problems and subtyping problems in addition to the binary tumor/normal classification tasks typically studied in weakly-supervised settings. By using attention-based learning, CLAM is also able to produce highly interpretable heatmaps that allow clinicians to visualize, for each slide, the relative contribution and importance of every tissue region to the model's predictions without using any pixel-level annotations during training. These heatmaps show that our models are capable of identifying well-known morphological features used by pathologists to make diagnostic determinations. Further, we show that the models are capable of distinguishing between tumor and adjacent normal tissue without any normal slides or ROIs used during training. CLAM is publicly available as an easy-to-use Python package over GitHub (https://github.com/mahmoodlab/CLAM), and whole slide-level attention maps can be viewed in our interactive demo (http://clam.mahmoodlab.org).
\end{spacing}
\begin{figure*}
\vspace{-12mm}
\includegraphics[width=\textwidth]{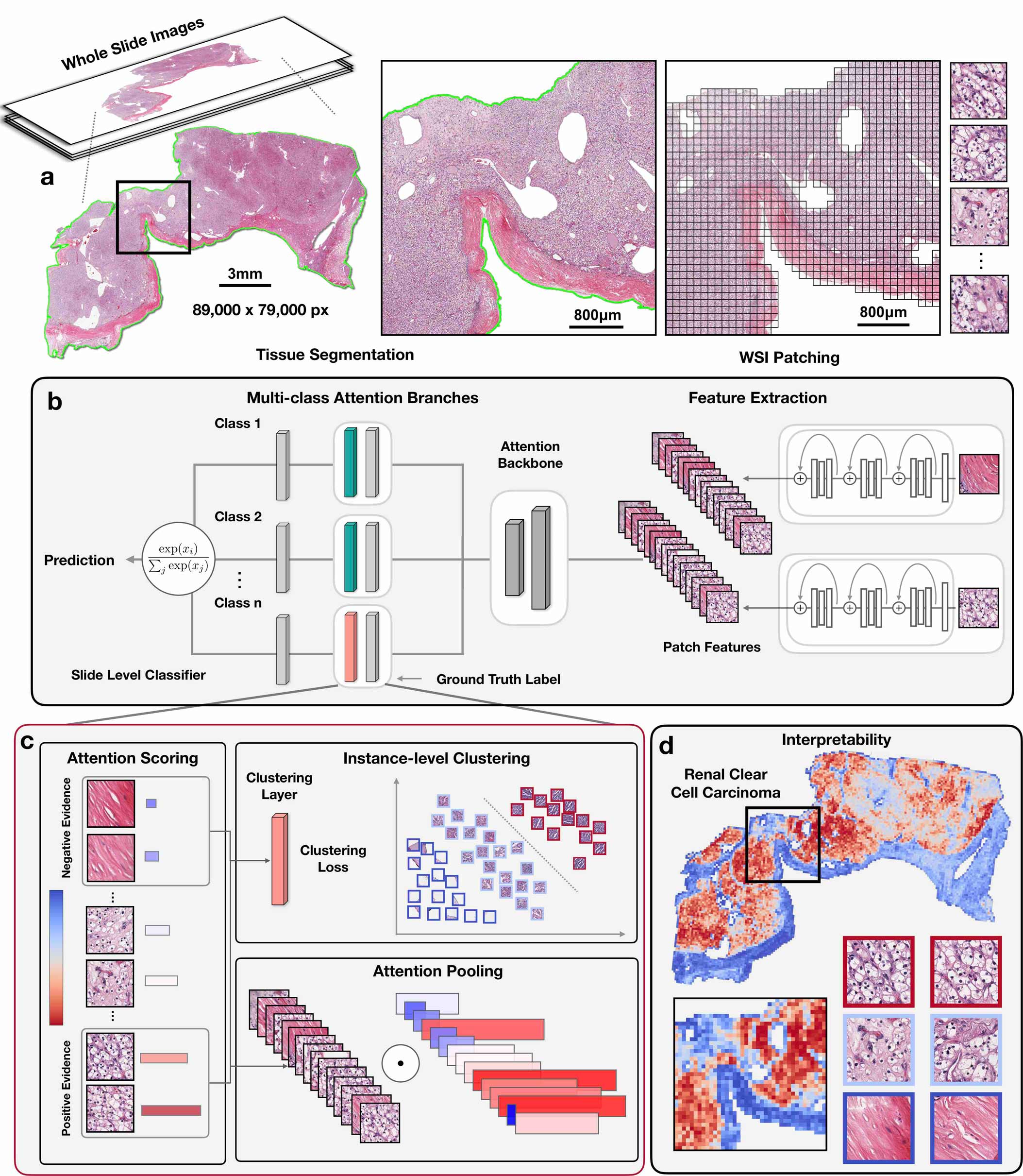}
\caption{\textbf{Overview of the CLAM conceptual framework, architecture and interpretability.} \textbf{a} Following segmentation, image patches are extracted from the tissue regions of the WSI. \textbf{b} Patches are encoded once by a pretrained CNN into a descriptive feature representation. During training and inference, extracted patches in each WSI are passed to a CLAM model as feature vectors. \textbf{c} For each class, the attention network ranks each region in the slide and assigns an attention score based on its relative importance to the slide-level diagnosis. Attention-pooling weighs patches by their respective attention scores and summarizes patch-level features into slide-level representations, which are used to make the final diagnostic prediction. The strongly attended and weakly attended regions are used as representative samples to train clustering layers that learn a rich patch-level feature space separable between the positive and negative evidence of distinct classes. \textbf{d} The attention scores can be visualized as a heatmap to identify ROIs and interpret the important morphology used for diagnosis.}
\end{figure*}
\begin{spacing}{1.45}
\vspace{-6mm}
\section*{CLAM: High-throughput, interpretable, weakly-supervised and data-efficient whole slide analysis}
\vspace{-4mm}
CLAM is a deep-learning-based weakly-supervised method that uses attention-based learning to automatically identify sub-regions of high diagnostic value in order to accurately classify the whole slide, while also utilizing instance-level clustering over the representative regions identified to constrain and refine the feature space.
For whole slide-level learning without annotation, CLAM uses an attention-based pooling function for aggregating patch-level features into slide-level representations for classification. At a high level, during both training and inference, the model examines and ranks all patches within the tissue regions of a WSI, assigning an attention score for each patch, which informs its contribution or importance to the collective, slide-level representation for a specific class (\textbf{Figure 1}). This interpretation of the attention score is reflected in the slide-level aggregation rule of attention-based pooling, which computes the slide-level representation as the average of all patches in the slide weighted by their respective attention score. Unlike the standard multiple instance learning algorithm\cite{maron1998framework,campanella2019clinical,ilse2018attention}, which was designed specifically for weakly-supervised positive/negative binary classification (\textit{e.g.} cancer vs. normal), CLAM is designed to solve generic multi-class classification problems. A CLAM model has $n$ parallel attention branches that together calculate $n$ unique slide-level representations, where each representation is determined from a different set of highly-attended regions in the image viewed by the network as strong positive evidence for the one of $n$ classes in a multi-class diagnostic task (\textbf{Figure 1-b, c}). Each class-specific slide representation is then examined by a classification layer to obtain the final probability score predictions for the whole slide.

\vspace{-5mm}
Under the MIL formulation and the weakly-supervised learning paradigm in general, one major bottleneck in developing high-performance machine learning classifiers for computational pathology is the inefficient usage of labeled WSI data. For example, when only the slide-level labels are known, despite having access to many (up to hundreds of thousands) instances or patches, the standard multiple learning algorithm uses max pooling and thus uses the gradient signal from only a single instance in each slide to update the learning parameters of the neural network model. This drawback can be used to partly explain why empirically, a deep learning model trained using MIL would require observing an enormous number of example WSIs annotated at the slide level to achieve high performance even for simple tasks such as binary classification.

\vspace{-5mm}
To address the data inefficiency in existing weakly-supervised learning algorithms for computational pathology, we adopt the attention-based pooling aggregation rule instead of max pooling. Additionally, we make use of the slide-level ground truth label and the attention scores predicted by the network to generate pseudo-labels for both highly attended and weakly attended patches as a novel means to increase the supervisory signals for learning a rich, separable, patch-level feature space. During training, the network learns from an additional supervised learning task of clustering the most and least attended patches of each class into distinct clusters. Additionally, it is possible to incorporate domain knowledge into the instance level clustering to add further supervision. For example, almost always, cancer subtypes are mutually exclusive; when one subtype is present in the WSI, once can assume that there is no morphology corresponding to other subtypes concurrently present. Following the mutual exclusivity assumption, in addition to supervising the attention branch for which the ground truth class is present, we supervise the attention network branches corresponding to the remaining classes by clustering their highly attended instances as "false positive" evidence for their respective classes.

\vspace{-5mm}
In order to make CLAM a high-throughput pipeline that researchers can readily adopt and utilize without requiring dedicated, high performance compute clusters, we also propose and make available an easy-to-use WSI processing toolbox. Our pipeline first automatically segments the tissue region of each slide and divides it into many smaller patches (\textit{e.g.} $256 \times 256$ pixels) so they can serve as direct inputs to a convolutional neural network (CNN) (\textbf{Figure 1a}). Next, using a pretrained CNN for feature extraction, we convert all tissue patches into sets of low-dimensional feature embeddings (\textbf{Figure 1b}). Following this feature extraction, both training and inference can occur in the low-dimensional feature space instead of the high-dimensional pixel space. The volume of the data space is decreased nearly 200-fold and we can drastically reduce the subsequent computation required to train supervised deep learning models. In practice, working with a low-dimensional feature space enables training models on thousands of gigapixel-sized resection slides within hours on modern consumer-grade workstations.
\vspace{-5mm}

In  proceeding sections, we demonstrate the data efficiency, adaptability and interpretability of CLAM on three different computational pathology problems: (A) renal cell carcinoma (RCC) subtyping (B) Non-small cell lung cancer (NSCLC) subtyping (C) Breast cancer lymph node metastasis detection. We additionally show that CLAM models trained on WSIs are adaptable to cell phone microscopy images and biopsy slides. 
\end{spacing}
\begin{spacing}{1.4}
\newpage
\section*{\large{Results}}
\vspace{-5mm}
\textbf{Dataset-Size Dependent, Cross-Validated Classification Performance.} \\
We evaluated the slide-level classification performance of CLAM for three clinical diagnostic tasks mentioned above using 10-fold monte carlo cross-validation. For each cross-validated fold, we randomly partitioned each public WSI dataset into a training set (80\% of cases), a validation set (10\% of cases) and test set (10\% of cases). When partitioning, the proportions of different classes (in terms of number of cases) are kept constant in each set. In the case that a single case has multiple slides, all of them are sampled into the same set. In each fold, the model's performance on the validation set is monitored during training and used for model selection while the test set is held-out and referred to just once after training is complete to evaluate the model. Models were additionally tested on independent test cohorts described in the next section. On the TGCA-Kidney dataset, at 20$\times$ magnification we achieved a 10-fold macro-averaged one-vs-rest mean test AUC of 0.991 (\textbf{Figure 2a}) for the three-class Renal Cell Carcinoma (RCC) subtyping of Papillary (PRCC), Chromophobe (CRCC) and Clear Cell Renal Cell Carcinoma (CCRCC). For the two-class Non-small Cell Lung Carcinoma (NSCLC) subtyping of Adenocarcinoma (LUAD) and Squamous Cell Carcinoma (LUSC) on the combined TCGA and CPTAC Lung dataset, at 20$\times$ magnification we achieved an average test AUC of 0.956 (\textbf{Figure 2b}). On the combined Camelyon16 and Camelyon17 dataset for breast cancer metastasis detection in axillary lymph nodes, at 40$\times$ magnification we achieved an average test AUC of 0.953 (\textbf{Figure 2c}). All of our training data come from publicly available sources which despite representing some of the largest public WSI datasets, are 5 to 10 times smaller than proprietary, labeled datasets studied by several recent works\cite{campanella2019clinical,bulten2020automated}. However, despite the moderate sizes of our datasets (884, 1967 and 899 total slides respectively, of which only around 80\% are used for training in each fold), the high performance ($>0.95$ AUC) on all three tasks indicate that our method can be effectively applied to solve both conventional positive vs. negative cancer detection binary classification and more general multi-class cancer subtyping problems across a variety of tissue types.

\begin{figure*}
\vspace{-8mm}
\includegraphics[width=\textwidth]{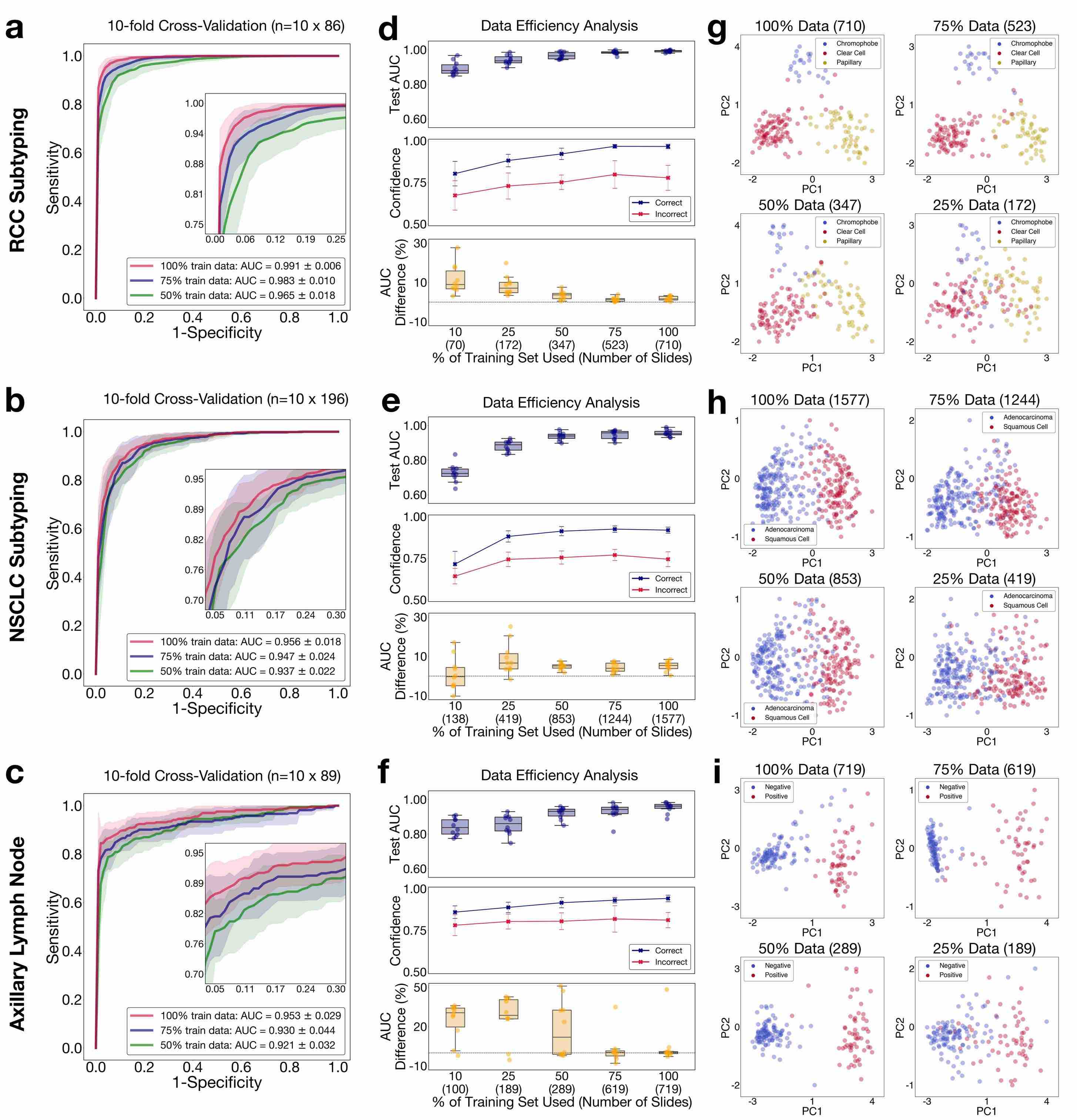}
\caption{\textbf{Performance, data efficiency and comparative analysis. a-c} The classification performance of CLAM models trained on slide-level labels using 100\%, 75\% and 50\% of cases in the training set is reported for each task (mean test AUC $\pm$ std from 10-fold monte-carlo cross-validation; confidence band shows $\pm$1 std for the averaged ROC curve). Using the entire training set, CLAM achieves a macro-averaged one-versus-rest test AUC of 0.991 for RCC subtyping, AUC of 0.956 for NSCLC subtyping and AUC of 0.953 for lymph node met. detection. \textbf{d-f} Top: the dataset-size-dependent performance of CLAM in terms of the 10-fold test AUCs for each training set size (100\%, 75\%, 50\%, 25\%, 10\% of cases) is shown using box plots (n=10). Middle: the mean confidence ($\pm$1 std) of the model's prediction for correctly and falsely classified slides. Bottom: box plots of the single-fold test AUC difference between CLAM and MIL (mMIL for RCC subtyping) show that CLAM performs better for nearly all training set sizes in all tasks. Dashed lines indicate the zero-difference level . \textbf{g-i} Trained CLAM models with common validation and test sets are used to visualize the separation of different classes in the learned slide-level feature space; the attention-pooled slide-level feature representation used for the model's prediction is plotted for each slide in the validation and test set following PCA.}
\end{figure*}
\vspace{-4mm}
Labeled WSI data are expensive and difficult to acquire, and it may not be feasible to collect over thousands of slides in the context of rare diseases (\textit{e.g.} Chromophobe RCC) or clinical trials. In light of this limitation, for each cross-validated fold created, we sequentially sampled subsets of training data equal to 75\%, 50\%, 25\% and 10\% of the total number of cases in the training set, while keeping the validation and test set the same, in order to investigate the dependency of the model's performance on the amount of training data available. When supervising CLAM models with the smaller, sampled subsets of training data, we observed that the number of slides required to achieve satisfactory performance (AUC$>0.9$) varies depending on the classification task. For example, merely 25\% of the total available training cases (which represents on average around 170 slides in each cross-validated fold) is sufficient to achieve an average test AUC of $>0.94$ on RCC subtyping. However, 50\% of the lung training set ($\sim$853 slides) and 50\% of the lymph node metastasis dataset ($\sim$289 slides) might be needed for NSLCC subtyping and lymph node metastasis detection respectively. Finally, we find in our comparative study, that CLAM consistently outperforms the max-pooling-based MIL algorithm (\textbf{Figure 2, d-f, bottom}) for all tasks and training set sizes (for full comparisons, see \textbf{Extended Data Figure 5} and \textbf{Supplementary Table 1-3}). Overall, we note that CLAM is surprisingly data efficient as it is often able to achieve test AUC $>0.9$ using only several hundred slides for training.
\end{spacing}


\begin{spacing}{1.5}
\vspace{-7mm}
\noindent\textbf{Generalization to Independent Test Cohorts.}
\begin{figure*}
\includegraphics[width=\textwidth]{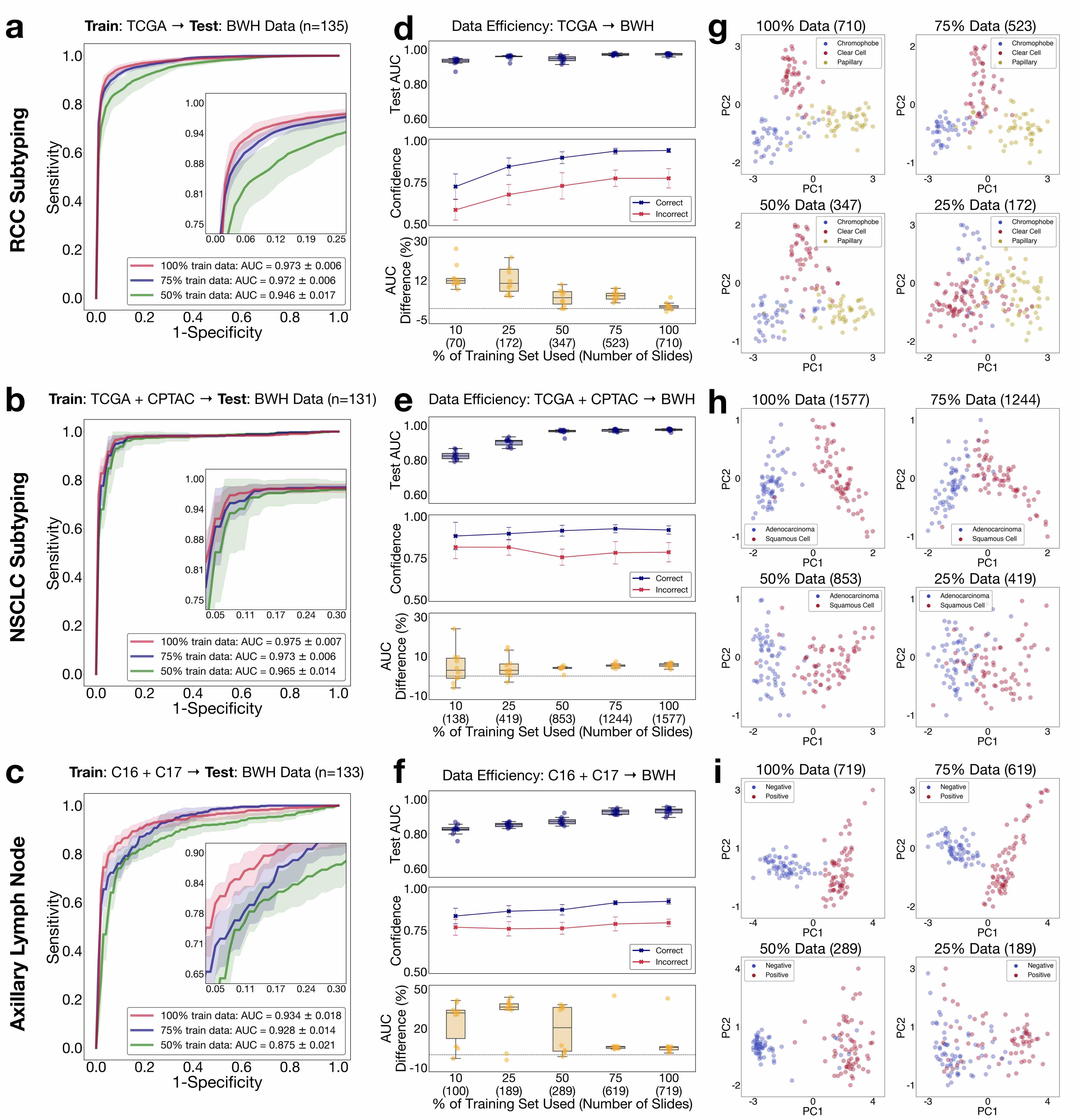}
\vspace{-8mm}
\caption{\textbf{Adaptability to independent test cohorts. a-c} Independent test cohorts collected at BWH are used to assess the capability of CLAM models trained on publicly available datasets to generalize to new data sources not encountered during training. Using 100\% of the training set, CLAM achieves a macro-averaged one-versus-rest AUC of 0.973 for kidney cancer subtyping (for per class AUC ROC, see \textbf{Extended Data Figure 4}), AUC of 0.975 for lung cancer subtyping and AUC of 0.934 for lymph node metastasis detection. \textbf{d-f} Data efficiency experiments demonstrate that CLAM models can generalize to unseen data despite having trained on limited number of labeled slides (top). The model makes higher confidence predictions for correctly classified than for incorrectly classified predictions and in general becomes less confident when trained using fewer data (middle). The difference in test AUC between CLAM and MIL (mMIL for RCC subtyping) is plotted for all 10 pairs of models in each training set size, showing CLAM outperforms mMIL/MIL (bottom) with high consistency. \textbf{g-i} For each task, visualization of the slide-level feature space for slides in the independent test cohorts using PCA reveals clear class separation diminishing with fewer training data.}
\end{figure*} \\
\noindent WSIs can vary greatly in image appearance due to differences in institutional standards and protocols for tissue processing, slide preparation and digitization. Therefore, it is important to validate that models trained under the CLAM weakly-supervised framework using moderate-sized, publicly available data sources are robust to data-source-specific variables and generalize to real-world clinical data from sources not encountered during training. We collected and scanned a total of 135 RCC (CRCC: 43, CCRCC: 46, PRCC: 46), 131 NSCLC (LUAD: 63, LUSC: 68), and 133 lymph node (Negative: 66, Positive: 67) whole slides at the Brigham and Women's Hospital as independent test cohorts for evaluating the generalization performance of our trained models (further explained in Methods and \textbf{Supplementary Table 11}). For each task and each training set size, the 10 models trained during cross-validation on our public datasets were directly evaluated on the completely held-out independent test set. We observed that for smaller denominations of the training set, the variance in the cross-validation performance of different models were often much higher, in which case testing using a single, best performing model may give the illusion of data-efficiency even though the algorithm's performance on the independent test set would be inconsistent and vary highly across models developed using different random splits of training data. To mitigate this, we used the average performance of all 10 models (instead of a single, selected model) to estimate the performance of our algorithm for each training set size (performance of the best cross-validation model on the independent test cohort with 95\% CI is available in Supplementary tables 13-15). When testing on independent test cohorts, the 10-fold cross-validated CLAM models trained using 100\% of the training set achieved an average one-vs-rest AUC (macro-averaged) of 0.973 on RCC subtyping, an average AUC of 0.975 on NSCLC subtyping, and average AUC of 0.934 on axillary lymph node metastasis detection respectively (\textbf{Figure 3, a-c}). Additionally, we observed that even CLAM models trained on the smaller subsets of the full training set can achieve respectable performance (test AUC $>0.9$) on data from independent sources after learning from just hundreds of slides (\textbf{Figure 3, d-f, top}).
When compared with mMIL/MIL, CLAM delivers improved performance across all tasks and training set sizes (\textbf{Figure 3, d-f, bottom}). For RCC subtyping and lymph node metastasis detection, CLAM outperformed mMIL and MIL respectively, especially when constrained by limited training data (\textit{e.g.} +14.5\% average test AUC for RCC subtyping and +30.1\% average test AUC for lymph node metastasis for models trained on 25\% of cases). Similarly, on the NSCLC subtyping task, CLAM outperformed MIL by on average 3-5\% in test AUC depending the training set size (for full comparisons, see \textbf{Extended Data Figure 5}). Additionally, We observed that CLAM models were more confident in their correct predictions than in their incorrect predictions. Furthermore, CLAM models became less confident as the size of the training set was reduced (\textbf{Figure 3, d-f, middle}), which is in general more desirable than having inaccurate yet overly confident models that severely and erroneously overfit on the small training set that they observe. Overall, the results from our study are highly encouraging and serve as supporting evidence that by using CLAM, moderately-sized datasets curated from multiple institutions (with source-specific variability) and a diverse patient distribution (\textit{e.g.} TCGA) are sufficient to develop accurate, weakly-supervised computer-aided diagnostic models capable of generalization. For best performance during real-world clinical deployment, we additionally propose to ensemble the diagnostic predictions from multiple models instead of selecting a single model (note this is easy and computationally inexpensive to do since we only have to perform feature extraction on our data once unlike other methods that require a unique feature encoder for each model). The ensemble performance (with 95\% CI) of trained CLAM models on all independent test cohorts is demonstrated in \textbf{Extended Data Figure 6} and \textbf{Supplementary Table 13-15}.
\end{spacing}
\begin{spacing}{1.40}
\vspace{-6mm}
\noindent\textbf{Interpretability and Whole Slide Attention Visualization.}
\begin{figure*}
\vspace{-12mm}
\includegraphics[width=\textwidth]{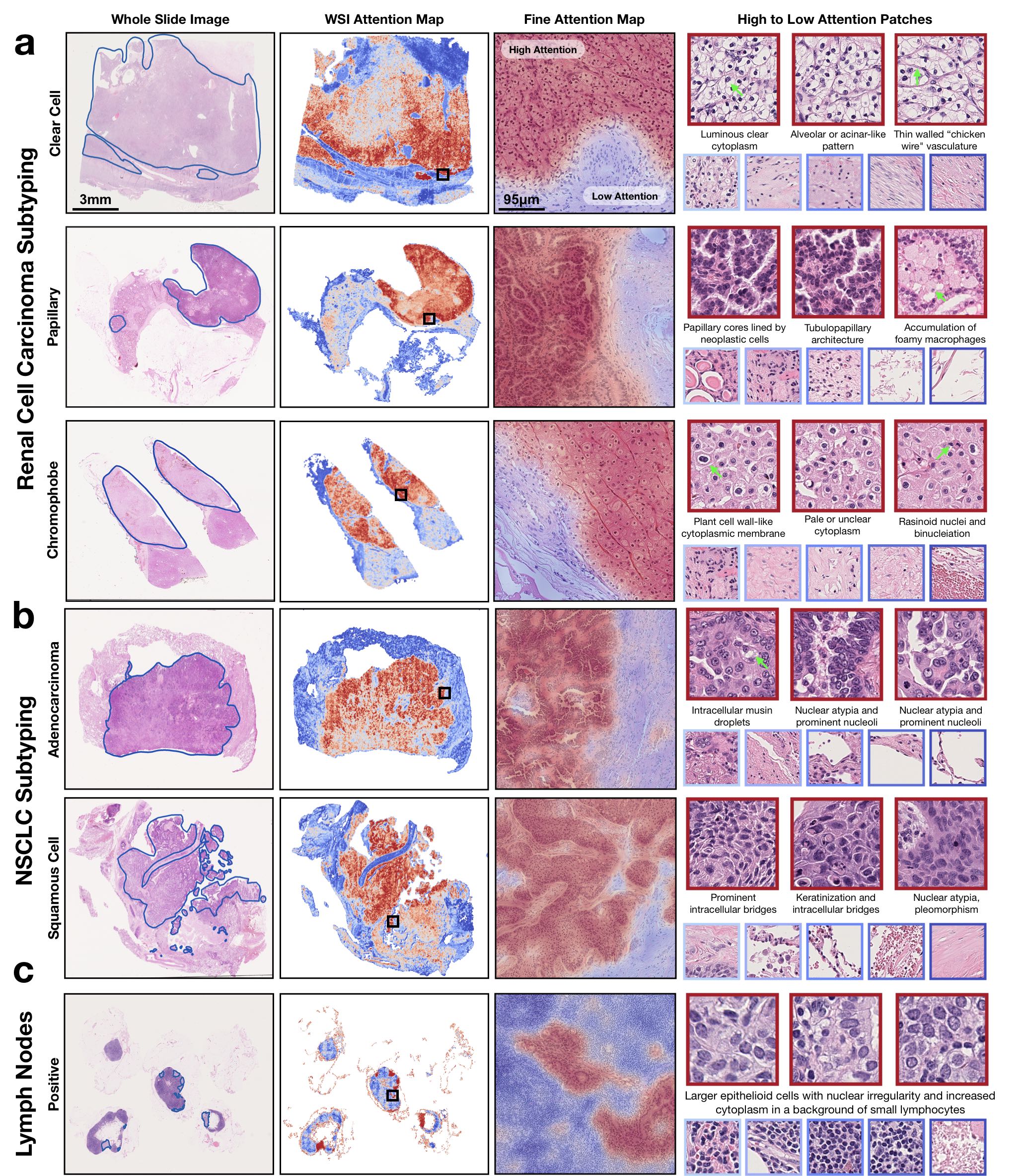}
\caption{\textbf{Interpretability and visualization.} For each RCC and NSCLC subtyping (\textbf{a, b}), a representative slide from each subtype is annotated by a pathologist (left), roughly highlighting the tumor tissue regions. Similarly, for a case of lymph node metastasis (\textbf{c}), regions of micro and macro-metastasis are highlighted. For each annotated slide, the whole slide attention heatmap is generated by computing the attention scores for the model's predicted class over patches tiled with a 25\% spatial overlap; the fine-grained ROI heatmap, which highlights parts of the smooth tumor normal boundary, is generated using a 95\% overlap and is overlaid onto the original H\&E image. Patches of the most highly attended regions (red border) generally exhibit well-known tumor morphology and low attention patches (blue border) include normal tissue among different background artifacts. Pathologist annotations are only for comparative analysis, no pixel level annotations were used for training. High resolution whole slide images and heatmaps corresponding these images may be viewed in the interactive demo accompanying this paper.  }
\end{figure*} \\
\noindent Human-readable interpretability of the trained weakly-supervised deep learning classifier can not only serve to validate that the predictive basis of the model aligns with well-known morphology used by pathologists for clinical diagnosis, but also has the potential to elucidate new morphological features of diagnostic relevance. A CLAM model makes its slide-level prediction by first identifying and aggregating regions in the WSI that are of high diagnostic importance (high attention score) while ignoring regions of low diagnostic relevance (low attention score). In order to visualize and interpret the relative importance of each region in the WSI, we can generate an attention heatmap by converting the attention scores for the model's predicted class into percentiles and mapping the normalized scores to their corresponding spatial location in the original slide. Fine-grained attention heatmaps can be created by using overlapping patches (\textit{e.g.} 95\% overlap) and averaging the attention scores in the overlapped regions (see \textbf{Extended Data Figure 3} for a discussion on the visual quality of heatmaps for different degrees of overlap). Even though pixel-level or patch-level annotation was never used during training to explicitly inform the model whether each region is tumor tissue (and if so, which subtype of tumor), we observe that through weakly-supervised learning using slide-level labels only, trained CLAM models are generally capable of identifying the boundary between tumor and normal tissue (\textbf{Figure 4 a-c}, see interactive demo at http://clam.mahmoodlab.org for high resolution heatmaps). This is an especially welcoming property since for RCC subtyping and NSCLC subtyping, all training data collected from the TCGA are positive cases and contain tumor regions. The finding demonstrates that CLAM has the potential to be used towards meaningful whole-slide-level interpretability and visualization in cancer subtyping problems for clinical or research purposes, without needing to observe negative cases during training (which would require either collecting slides from adjacent normal tissue or manual annotation of negative regions in positive slides). Of equal importance, for all three classification tasks studied, high attention regions generally correspond with morphology already established and recognized by pathologists (\textbf{Figure 4 a-c}). For example, the CLAM model trained for NSCLC subtyping highlights prominent intracellular bridges and keratinization and uses them as strong evidence (high attention) for Squamous Cell Carcinoma (\textbf{Figure 4b}), in concordance with human pathology expertise. For a positive case of lymph node metastasis, we additionally examined the model's attention heatmap with corresponding cytokeratin (AE1/AE3) immunohistochemical (IHC) staining to further validate its predictive basis (\textbf{Extended Data Figure 2}). However, caution should be taken not to overly rely on the attention heatmaps with the expectation that they can serve as pixel-perfect segmentation masks; intuitively, the attention scores for each region in the slide are relative and simply represent the model's interpretation of which regions are more important (relative to others) in determining the slide-level prediction. Nonetheless, armed with this simple, intuitive yet powerful interpretability and visualization technique, researchers can potentially look to discover previously unknown biomarkers for diseases by identifying highly attended regions that share relatively consistent, conserved morphological patterns across cases. \\\noindent\textit{Further Interpretability by visualizing the patch-level feature space.}
As a means to further investigate the patch-level feature space learned by CLAM models, we randomly sampled 2\% of patches from each slide in the independent test cohorts (a total of 54,995 patches from the RCC dataset, 64,687 patches from the NSCLC dataset, and 136,726 patches from the lymph node metastasis dataset) and applied PCA to reduce the 512-dimensional patch-level feature space learned by the models to two dimensions. For RCC and NSCLC subtyping (\textbf{Extended Data Figure 1, a-b}), each patch is colored with the class for which the corresponding instance-level clustering network layer predicted positive ($p  \geq 0.5$). Patches predicted as negative ($p < 0.5$) for every single class are labeled as "Agnostic". Patches of different predicted classes are separated into distinct clusters in the feature space and exhibit morphology characteristic of their respective subtype. For axillary lymph node metastasis detection, sampled positive patches (predicted as positive with $p  \geq 0.5$) include tumor tissue while agnostic patches ($p < 0.5$) capture a wide array morphology including healthy tissue and dense immune cell populations (\textbf{Extended Data Figure 1c}). 

\end{spacing}

\begin{spacing}{1.4}
\vspace{-6mm}
\noindent\textbf{Adapting networks trained on whole slide images to cellphone microscopy images.}
\begin{figure*}
\vspace{-5mm}
\includegraphics[width=\textwidth]{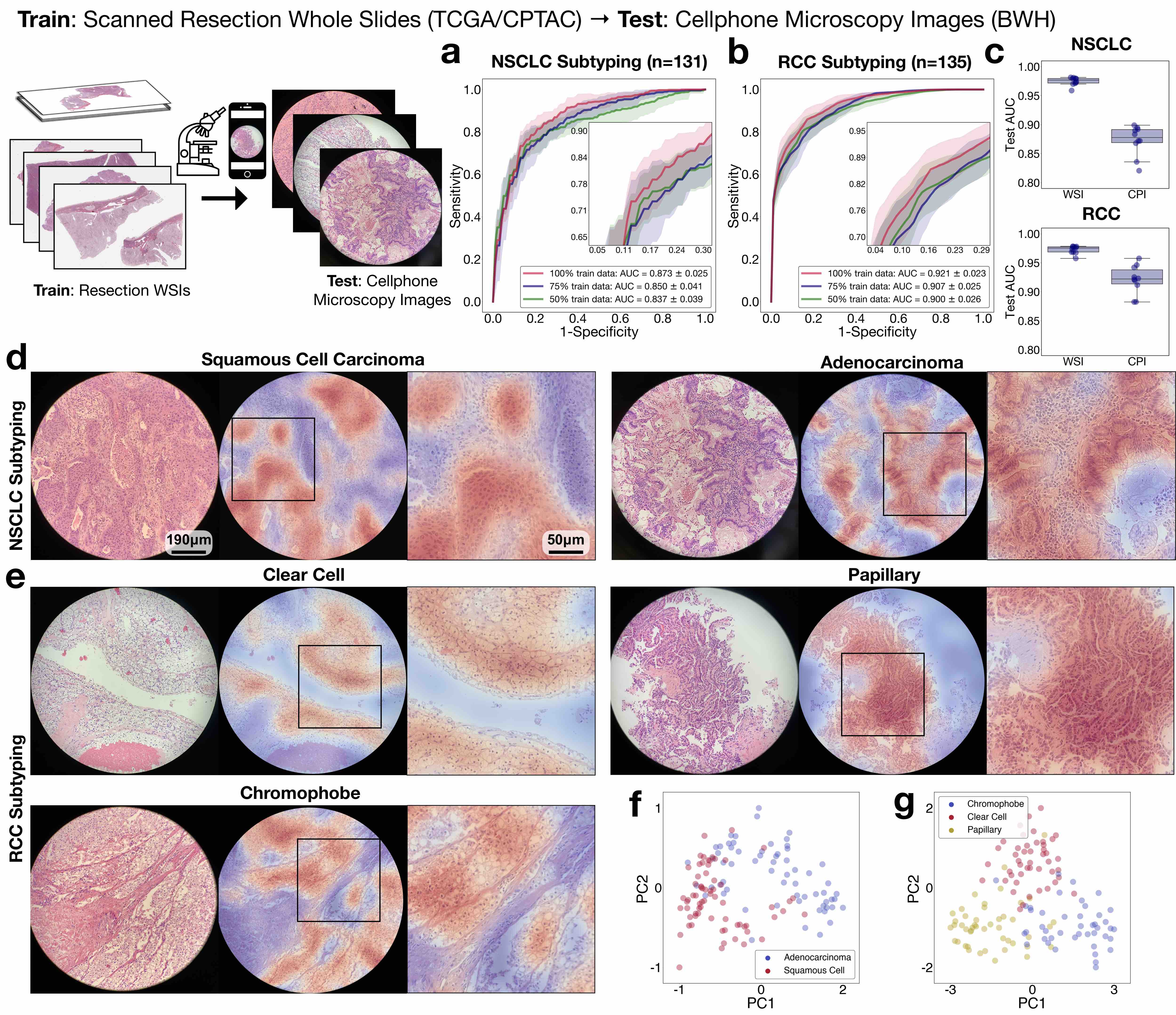}
\caption{\textbf{Adaptability to cellphone microscopy images.} \textbf{a, b} CLAM models trained on WSIs can adapt to images taken with a consumer-grade cellphone camera (CPIs) without further fine-tuning, achieving an average test AUC of 0.873 and 0.912 on the BWH NSCLC and RCC independent datasets respectively. For each slide, 4-8 representative fields-of-view (FOVs) are selected and captured at 20$\times$ magnification by a pathologist while viewing through the microscope. For each slide, patches extracted from all FOVs are collectively used by the CLAM model to inform the slide-level diagnosis. \textbf{c} A drop in performance is expected when directly adapting models trained on data from one imaging modality (WSIs) to another (CPIs). We noted a decrease of 0.102 and 0.051 in mean test AUC (relative to the performances on the corresponding WSI independent datasets) for NSCLC and RCC subtyping respectively when evaluating CLAM models (using 100\% of training set) on our CPI datasets. \textbf{d, e} The attention heatmap helps make the model's prediction interpretable by highlighting in each FOV, the discriminative regions used by the model to make the slide-level diagnostic prediction. We observed that the model attends strongly to tumor regions and largely ignores normal tissue and background artifacts as expected. However, due to the circular-shaped cutout of each FOV, patches near the border inevitably encapsulates varying degree of black space in addition to the tissue content, which can mislead the model toward assigning weaker attention to those regions than otherwise. \textbf{f, g} As an additional validation that CLAM models trained on WSIs are directly applicable to the classification of CPIs, we visualized the attention-pooled feature representation of each set of CPIs and observed that there is visible separation between distinct classes in both the NSCLC and RCC cellphone datasets.}
\end{figure*} \\
\noindent We additionally explored the ability of our models (which are trained exclusively on  WSIs) to directly adapt to microscopy images captured using a cellphone camera. In resource-constrained areas with limited access to pathologist expertise, consult cases are often imaged by cellphone. Training a deep learning classifier specifically based on cellphone microscopy images would likely require the time consuming, laborious process of manually curating a large set of labeled ROIs. These ROIs should not only be representative of the underlying pathological conditions but also capture a wide range of tissue-site and patient-specific appearances and artifacts to ensure the model can adapt to the heterogeneity inherently found in Histopathlogy slides and WSIs. Therefore, a robust model trained on WSIs that is capable of directly adapting to cellphone images (CPIs) and deliver accurate automated diagnosis is therefore of tremendous value to the wider adoption of telepathology. As part of our model adaptability study, 4-8 fields-of-view (FOVs) from each slide in our independent test cohorts are captured using a consumer-grade smartphone camera and patches from all FOV ROIs are collectively used by the model to predict the slide-level label. On the lung CPI dataset, CLAM achieved an average test AUC of 0.873 with the best model scoring an AUC of 0.899 (\textbf{Figure 5a}). On the kidney CPI dataset, CLAM achieved an average one-vs-rest macro-averaged AUC 0.921 with the best model scoring an AUC of 0.958 (\textbf{Figure 5, a-b} and \textbf{Supplementary Table 7-8}). The drop in performance compared to testing on WSIs (\textbf{Figure 5c}) can likely be attributed to the imperfect conditions under which cellphone images are captured (poor focus, non-uniform illumination, noise artifacts, vignetting, color shift, magnification changes, \textit{etc.}). While some of these adversities can potentially be reduced by using both conventional and deep learning-based image processing techniques (\textit{e.g.} stain normalization based on deep convolutional adversarial generative modeling\cite{bentaieb2017adversarial}), we did not attempt to correct or normalize the images in order to test the robustness and adaptability of our models and to keep the processing time and computational cost low to potentially allow inference directly on cellphone hardware. Surprisingly, despite these challenging variables, we find that in most cases, the model still accurately attends to regions in the FOV that exhibit well-known morphology characteristic of each cancer subtype (\textbf{Figure 5, d-e}). Further, distinct classes are still visibly separated into distinct clusters in the feature space that the model has learned from WSIs (\textbf{Figure 5, f-g}). These results instill confidence with regards to the potential wider applicability of our weakly-supervised learning framework to the telepathology domain.
\end{spacing}
\vspace{-5mm}
\noindent\textbf{Adapting networks trained on resections to biopsies.} 
\vspace{-5mm}
\begin{spacing}{1.38}
\begin{figure*}
\includegraphics[width=\textwidth]{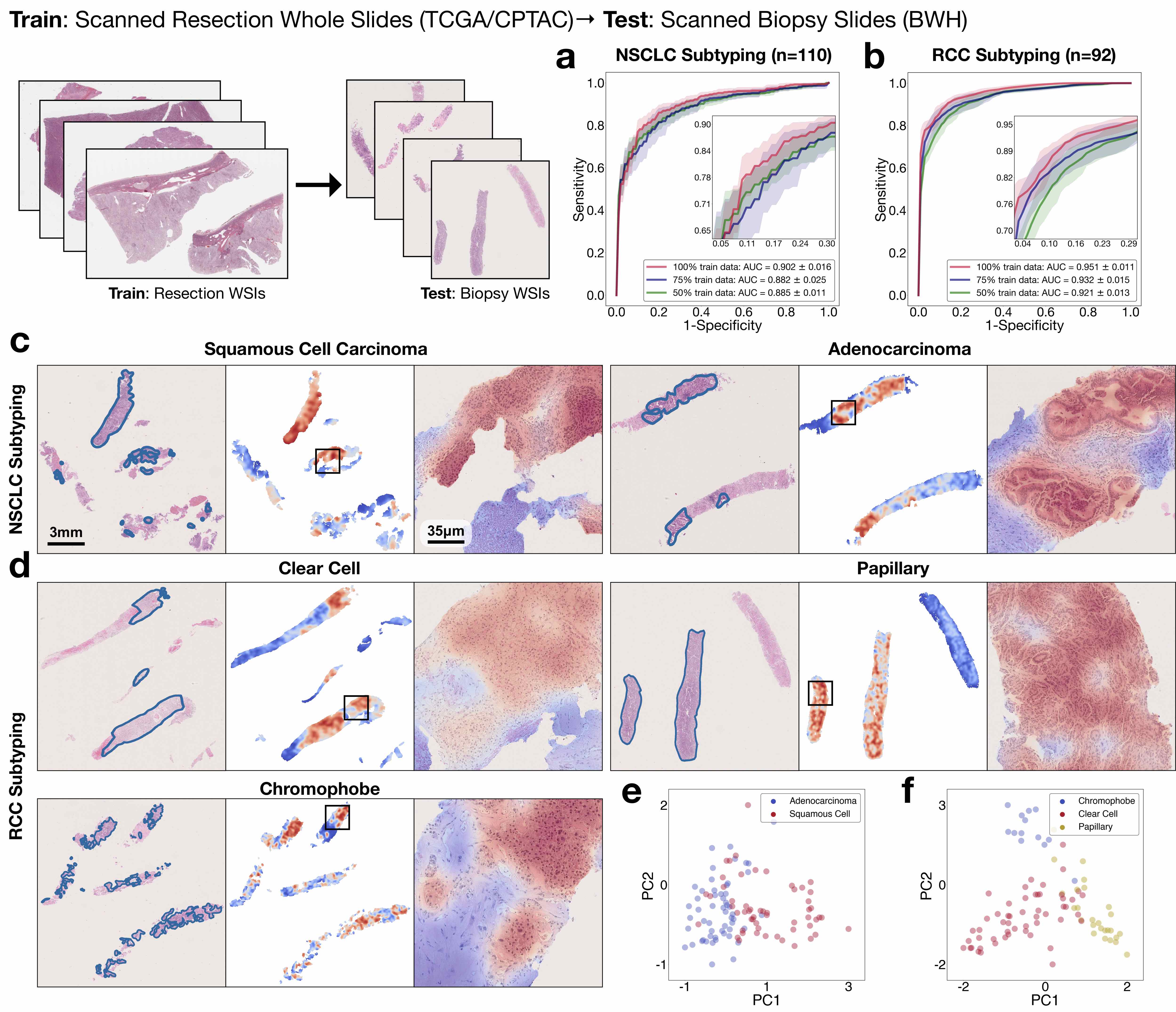}
\caption{\textbf{Adaptability to biopsy slides.} Compared with resection WSIs, biopsy WSIs generally contain much lower tissue content (average number of patches extracted from the tissue regions of each slide is 820 in our BWH lung biopsy dataset compared to 24714 in the lung resection dataset). The presence of crush artifacts and poorly differentiated and sparsely distributed tumor cells can further challenge accurate diagnosis. \textbf{a, b} We observed that CLAM models trained on resections are directly adaptable to biopsy WSIs, achieving a respectable average test AUC of 0.902 and 0.951 respectively on our NSCLC and RCC biopsy independent test cohorts without further fine-tuning or ROI extraction. \textbf{c, d} Attention heatmap visualization for biopsy slide (left: H\&E slide with pathologist annotation for tumor regions. mid: heatmap for patches tiled with 95\% overlap. right: zoomed-in view of tumor regions attended by the CLAM model) is consistent with our findings on the resection and cellphone datasets as the regions most strongly attended by the model consistently correspond to tumor tissue. The attention heatmaps also tend to clearly highlight the tumor/normal tissue boundaries even though no patch-level or pixel-level annotation was required or used during training. \textbf{e, f} The slide-level feature representations of the biopsy datasets are visualized using PCA. We observed that for both RCC and NSCLC, the feature space learned by the CLAM model from resections remains visibly separable among the distinct subtypes when it is adapted to biopsy slides. A high resolution version of these biopsy whole slides and heatmaps may be viewed in the interactive demo accompanying this paper.}
\end{figure*}
\noindent The publicly available WSIs that we used for training in our study are all resections. Compared to resected tissue, core needle biopsied tissue are generally significantly smaller in size. The limited tissue content and the presence of cell distortion due to crush artifact can challenge the model's diagnostic ability. Accordingly, since we did not use any biopsy slides during training, it was important to investigate whether models trained on solely resections can adapt directly to biopsy slides and make accurate diagnostic predictions. We collected 110 lung (55 LUAD and 55 LUSC) and 92 kidney biopsy slides (53 CCRCC, 26 PRCC, 13 CRCC) at BWH as our independent test cohorts and directly tested our models that had been trained on the publicly available resection data. On the lung biopsy test set, CLAM achieved an average AUC of 0.902 with the best model scoring an AUC of 0.926. On the kidney biopsy test set, the average one-vs-rest macro-averaged test AUC is 0.951 with the best model scoring an AUC of 0.967 (\textbf{Figure 6, a-b} and \textbf{Supplementary Table 9-10}). These results are highly encouraging because many biopsy slides, especially in the case of the lung biopsy dataset, contained poorly differentiated tumors that make them highly difficult or impossible for pathologists to accurately diagnose based on the H\&E stains alone (without IHC). Additionally, in order to assess the applicability of our models to potential real-world, fully-automated computer aided diagnosis, when testing on biopsy slides, we did not manually select ROIs that contain high tumor content as a means to avoid exposing the model to non-tumor features (blood vessels, inflammation, necrotic regions, \textit{etc.})\cite{natmedlung} that might lead to misclassification. We also did not perform any preprocessing techniques such as stain normalization on our test set and used the entire tissue region of each slide during evaluation. Using the same visualization and interpretability technique as before, we generated attention heatmaps for one slide of each subtype (\textbf{Figure 6, c-d}). We continued to observe high similarity between the strongly attended regions highlighted by the trained CLAM models and the pathologist's annotated tumor regions despite the tumors generally occupying smaller and more sparse tissue regions than in resection slides.
\vspace{-8mm}
\section*{\large{Discussion}}
\vspace{-5mm}
We showed that CLAM addresses five key challenges in computational pathology: a) \textbf{Weak supervision:} We trained models with weak supervision using only slide-level labels without pixel-, patch- or ROI-level annotation. Weak supervision also objectively identifies relevant morphological features from the tissue microenviornment without any \textit{a priori} knowledge or subjective annotation. In three separate analyses, we showed that our models can identify well-known morphological features and accordingly, has the capability of identifying new morphological features of diagnostic, prognostic, and therapeutic relevance. b) \textbf{Data efficiency:} Our quantitative experiments and comparative analysis demonstrate that CLAM is more data efficient as compared to the established weakly-supervised whole slide classification approaches using multiple instance learning (Extended Data Figure 5). This reduces the trade off between weak supervision and number of labeled whole slides required for training. Data-efficient whole slide training also enables the applicability of computational pathology for classification in rare conditions, and patient stratification for clinical trials where it is valuable to predict response or resistance to treatment from a small cohort of existing patient cases. c) \textbf{Applicability to multi-class subtyping problems:} Unlike other computational pathology studies that focus on the application of weakly-supervised methods to computer-aided diagnosis in the binary (positive/negative) setting, we demonstrate that CLAM is applicable to multi-class subtyping problems and may be used to stratify patients between predominant and relatively rare classes (for example Clear Cell vs Chromophobe in Renal Cell Carcinoma). d) \textbf{Adaptability:} We demonstrate that models trained using CLAM are highly adaptable to independent test cohorts, biopsy slides and cellphone microscopy images. We demonstrate that models trained on histology whole slide images with relatively large tissue content are adaptable to data from biologically independent test cohorts of resection and biopsy slides. Training on resections (average tissue coverage: 142 $\text{mm}^{2}$, 11182 patches) and adapting to biopsies (average tissue coverage: 15.6 $\text{mm}^{2}$, 1225 patches) allows us to use limited slides while maximizing tissue content available for training. Furthermore, resource-constrained settings with limited anatomic pathology expertise often send consult cases imaged via a consumer-grade cellphone attached to brightfield microscopes. We showed CLAM models trained on whole slide images covering large tissue volume to cellphone images which have a limited field-of-view. e) \textbf{Interpretability and explainability as a clinical and research tool:} We demonstrate that our models are interpretable and capable of generating fine heatmaps which can isolate the tumor regions without using pixel-level annotation for identifying regions containing tumor. While no normal tissue or ROI extraction from adjacent normal tissue regions was used for training, our method is capable of identifying the most relevant regions used to make classification determinations. At this time, we have built a website hosting interactive demos for high resolution ($10\times$ magnification), finegrained heatmaps for at least 15 WSIs in our independent test sets (both biopsy and resection) that include all subtypes of cancer and tissue sites in our study. To the best of our knowlege, we are the first to showcase interpretability heatmap for the entire WSI at such high resolution and fine scale as proof of concept for a  computer-aided diagnostic system trained only with weak supervision. These heatmaps may be used as an interpretability tool in research applications to identify new morphological features associated with response and resistance to treatment or used as a visualization tool for secondary opinion in anatomic pathology. 
\end{spacing}
\vspace{-5mm}
\section*{\large{Online Methods}}
\begin{spacing}{1.5}
\vspace{-4mm}
\noindent\textbf{CLAM} \\
Clustering-constrained Attention Multiple Instance Learning (CLAM) is a high-throughput deep learning empowered toolbox designed to solve weakly-supervised classification tasks in computational pathology in which each whole slide image (WSI) in the training set is a single data point with known slide-level diagnosis but no class specific information or annotation is available for any pixel or region in the slide. CLAM builds upon the multiple instance learning framework which views each WSI (known as a bag) as comprising of many (up to hundreds of thousands) smaller regions or patches (known as instances). The original MIL algorithm restricted its scope to binary classification problems of a positive and a negative class based on the assumption that if at least one patch belongs to the positive class, then the entire slide should be classified as positive, whereas a slide should be classified as negative if all patches are of the negative class. This assumption is reflected in the rigid, non-trainable aggregation function of max pooling which simply uses the patch with the highest predicted probability for the positive class for the slide-level prediction, rendering MIL suitable for neither multi-class classification nor binary classification problems in which no intrinsic positive/negative assumption can be made. In contrast, CLAM is applicable to general multi-class classification as instead of using max pooling or other aggregation functions (\textit{e.g.} mean operator, generalized mean, log-sum-exp, the quantile function, Noisy-OR, Nosiy-And\cite{couture2018multiple, kraus2016classifying, noisy-or}, \textit{etc.}) of limited flexibility for dataset or problem-specific tuning, we build CLAM around the trainable, attention-based pooling function\cite{ilse2018attention} to aggregate slide-level representations from patch-level representations for each class. In our design of multi-class attention pooling, the attention network predicts $n$ distinct sets of attention scores corresponding to the $n$ classes in a multi-class classification problem. This enables the network to unambiguously learn for each class, which morphological features should be considered as positive evidence (characteristic of the class) vs. negative evidence (non-informative) and summarize $n$ unique slide-level representations. 
In CLAM, the first fully-connected layer $\mathbf{W}_{1} \in \mathbbm{R}^{512 \times 1024}$ further compresses each fixed 1024-dimensional patch-level representation $\mathbf{z}_{k}$ to a 512-dimensional vector $\mathbf{h}_{k}=\mathbf{W}_{1}\mathbf{z}_{k}^{\top}$ (for simplicity, all bias terms are implied and not explicitly written). If we consider the first two layers of the attention network $\mathbf{U}_{a} \in \mathbbm{R}^{256 \times 512} \text{ and } \mathbf{V_a} \in \mathbbm{R}^{256 \times 512}$ collectively as the attention backbone shared by all classes, the attention network then splits into $n$ parallel attention branches $\mathbf{W}_{a,1}, ... \thinspace, \mathbf{W}_{a,n} \in \mathbbm{R}^{1 \times 256}$. Similarly, $n$ parallel, independent classifiers, $\mathbf{W}_{c,1}, ... \thinspace \mathbf{W}_{c,n}$ are built to score each class-specific slide-level representation. Accordingly, the attention score of the $k^{\text{th}}$ patch for the $m^{\text{th}}$ class, denoted $a_{k,m}$, is given by eqn \ref{clsattention} and the slide-level representation aggregated per the attention score distribution for the $m^{\text{th}}$ class, denoted $\mathbf{h}_{slide,m} \in \mathbbm{R}^{1 \times 512}$, is given by eqn \ref{clsweightedaverage}:
\begin{equation} \label{clsattention}
a_{k,m}=\frac{\exp \left\{\mathbf{W}_{a,m}\left(\tanh \left(\mathbf{V}_{a} \mathbf{h}_{k}^{\top}\right) \odot \operatorname{sigm}\left(\mathbf{U}_{a} \mathbf{h}_{k}^{\top}\right)\right)\right\}}{\sum_{j=1}^{N} \exp \left\{\mathbf{W}_{a,m} \left(\tanh \left(\mathbf{V}_{a} \mathbf{h}_{j}^{\top}\right) \odot \operatorname{sigm}\left(\mathbf{U}_{a}\mathbf{h}_{j}^{\top}\right)\right)\right\}}
\end{equation}
\vspace{-4mm}
\begin{equation} \label{clsweightedaverage}
\mathbf{h}_{slide,m}=\sum_{k=1}^{N} a_{k,m} \mathbf{h}_{k}
\end{equation} 
The corresponding unnormalized slide-level score $s_{slide,m}$ is given via the classifier layer $\mathbf{W}_{c,m} \in \mathbbm{R}^{1 \times 512}$ by $s_{slide,m}=\mathbf{W}_{c,m} \mathbf{h}_{slide,m}^{\top}$. For inference, the predicted probability distribution over each class is computed by applying a softmax function to the slide-level prediction scores.
$\mathbf{s}_{slide}$. \\
\textit{Instance-level Clustering.}
To further encourage the learning of class-specific features, we introduce an additional binary clustering objective during training. For each of $n$ classes, we place a clustering layer with 512 hidden units after the first fully-connected layer, $\mathbf{W}_{1}$. If we denote the weights of the clustering network that corresponds to the $m^{th}$ class as $\mathbf{W}_{inst, m} \in \mathbbm{R}^{2 \times 512}$, the cluster assignment scores predicted for the $k^{th}$ patch, denoted by $\mathbf{p}_{m,k}$, is given as:
\begin{equation} \label{cluster}
\mathbf{p}_{m,k} = \mathbf{W}_{inst, m}\mathbf{h}_{k}^{\top}   
\end{equation}
Given we do not have access to patch-level labels, we use the outputs of the attention network to generate pseudo labels in for each slide in each iteration of training to supervise the clustering. Instead of clustering all the patches in the slide, we only optimize the objective over the subset of patches for which the model either strongly attends to or nearly completely ignores. Let the entire label set be $\mathcal{Y}=\{0, ..., \thinspace n-1\}$, to avoid confusion, for a given slide, with ground truth class label $Y \in \mathcal{Y}$, we refer to the attention branch that corresponds with this ground truth class ($\mathbf{W}_{a,Y}$) as "in-the-class", and the remaining $n-1$ attention branches as "out-of-the-class". If we denote the sorted list of in-the-class attention scores (in ascending order) as $\tilde{a}_{1,Y},...\thinspace,\tilde{a}_{N,Y}$, we take the $B$ patches with the lowest attention scores and assign them the negative cluster label ($y_{Y,b}=0, 1 \leq b \leq B$) while the $B$ patches with the highest in-the-class attention scores receive the positive cluster label ($y_{Y,b}=1, B+1 \leq b \leq 2B$). 
Intuitively, because each attention branch is supervised by the slide-level label during training, the $B$ patches with high attention scores (hence the positive cluster) are expected to be strong positive evidence for class $Y$, while the $B$ patches with low attention scores (hence the negative cluster) are expected to be strong negative evidence for class $Y$. Therefore, the clustering task can be intuitively interpreted as constraining the patch-level feature space $\mathbf{h}_{k}$ such that the strong characterizing evidence of each class is linearly separable from its negative evidence.
For cancer subtyping problems, often it is reasonable to assume that the presence of representative tissue patches from different classes are mutually exclusive events (\textit{i.e.} they cannot be present in the same slide), as we cluster the most attended and least attended patches of the in-the-class attention branch into positive and negative evidence respectively, it makes sense to also impose additional supervision on the $n - 1$ out-of-the-class attention branches. Namely, given the ground truth slide label $Y$, $\forall m \in \mathcal{Y}\backslash\{Y\}$,
the $B$ patches with the highest attention scores cannot be positive evidence for class $m$ provided that we assume none of the patches on the slide is of class $m$ (due to the mutual exclusivity). As a result, in addition to clustering the $2B$ patches selected from the in-the-class attention branch, we assign the negative cluster label to the top $B$ attended patches in all out-of-the-class attention branches since they are assumed to be “false positive” evidence. On the other hand, if the mutual exclusivity assumption does not hold (\textit{e.g.} cancer vs. no cancer problem, where a slide can contain both patches representative of tumor tissue and normal tissue), then we do not supervise the clustering of highly attended patches from out-of-the-class branches since we do not know if they are false positives or not. Using the aforementioned notations, the full instance-level clustering algorithm is summarized below in \textbf{Algorithm \ref{algo1}}.
\end{spacing}
\begin{singlespace}
\begin{algorithm}[h!] 
    \caption{Instance-level Clustering} \label{algo1}
    \begin{algorithmic}
        \Function{Cluster}{$(\mathbf{h}_1, \mathbf{a}_1),... \thinspace,(\mathbf{h}_K, \mathbf{a}_K), Y$}
            \For{$m \gets 1,2,...\thinspace, n$} 
                \If{$m = Y$}
                    \State $(\mathbf{\tilde{h}}_{1}, \tilde{a}_{1,m}),... \thinspace,(\mathbf{\tilde{h}}_{K},\tilde{a}_{K,m})=\underset{a_{k,m}}{\textbf{SortAscending}}((\mathbf{h}_{1}, a_{1,m}),...\thinspace, (\mathbf{h}_{k}, a_{k,m}),...\thinspace,(\mathbf{h}_{K}, a_{K,m}))$
                    \For{$b \gets 1,...,B$}
                        \State \textit{\{generate pseudo label for positive and negative evidence\}}
                        \State $y_{m,b}=0$ \Comment{negative evidence}
                        \State $y_{m,b+B}=1$ \Comment{positive evidence}
                        \State \textit{\{cluster assignment prediction\}}
                        \State $\mathbf{p}_{m,b} = \mathbf{W}_{inst, m}\mathbf{\tilde{h}}_{b}^{\top}$ \Comment{prediction for negative evidence}
                        \State $\mathbf{p}_{m,b+B} = \mathbf{W}_{inst, m}\mathbf{\tilde{h}}_{K-B+b}^{\top}$ \Comment{prediction for positive evidence}
                    \EndFor
                \Else
                    \If{classes are mutually exclusive}
                        \State $(\mathbf{\tilde{h}}_{1}, \tilde{a}_{1,m}),... \thinspace,(\mathbf{\tilde{h}}_{K},\tilde{a}_{K,m})=\underset{a_{k,m}}{\textbf{SortAscending}}((\mathbf{h}_{1}, a_{1,m}),...\thinspace, (\mathbf{h}_{k}, a_{k,m}),...\thinspace,(\mathbf{h}_{K}, a_{K,m}))$
                        \For{$b \gets 1,...,B$}
                            \State \textit{\{generate pseudo label for false positive evidence\}}
                            \State $y_{m,b}=0$ \Comment{false positive evidence}
                            \State \textit{\{cluster assignment prediction\}}
                            \State $\mathbf{p}_{m,b} = \mathbf{W}_{inst, m}\mathbf{\tilde{h}}_{K-B+b}^{\top}$ \Comment{prediction for false positive evidence}
                    \EndFor
                    \Else
                        \State \textbf{pass}
                    \EndIf
                \EndIf
            \EndFor
            \If {classes are mutually exclusive}
                \State \Return{$[\mathbf{p}_{1},...\thinspace,\mathbf{p}_{n}],\thinspace[\mathbf{y}_{1},...\thinspace,\mathbf{y}_{n}]$}
            \Else
                \State \Return{$[\mathbf{p}_{Y}],\thinspace[\mathbf{y}_{Y}]$}
            \EndIf
            
        \EndFunction
\end{algorithmic}
\end{algorithm}
\end{singlespace} 
\begin{spacing}{1.45}
\vspace{-4mm}
\noindent\textit{Smooth SVM Loss.}
For the instance-level clustering task, we chose to use the smooth top$1$ SVM loss\cite{topk}, which is based on the well-established multiclass SVM loss\cite{multiclasssvm}. In a general $n$-class classification problem, neural network models output a vector of prediction scores $\mathbf{s}$, where each entry in $\mathbf{s}$ corresponds to the model's prediction for a single class. Given the set of all possible ground truth labels $\mathcal{Y} = \left\{0,1,...,n-1\right\}$ and ground truth label $y \in \mathcal{Y}$, the multiclass SVM loss penalizes the classifier linearly in the difference between the prediction score for the ground truth class and the highest prediction score for remaining classes only if that difference is greater than a specified margin $\alpha$ (eqn \ref{svmloss}). The smoothed variant (eqn \ref{smoothsvmloss}) adds a temperature scaling $\tau$ to the multiclass SVM loss, with which it has been shown to be infinitely differentiable with non-sparse gradients and suitable for the optimization of deep neural networks when the algorithm is implemented efficiently. The smooth SVM loss can be viewed as a generalization of the widely used cross-entropy classification loss for different choices of finite values for the margin and different temperature scaling. \\
The introduction of a margin to the loss function has been empirically shown to reduce over-fitting when the data labels are noisy or when data are limited. During training, the pseudo labels we create to supervise the instance-level clustering task are expected to be noisy. Namely, the top attended patches might not necessarily correspond to the ground truth class and likewise, the least attended patches are also not guaranteed to be actual negative evidence of the class. Therefore, instead of the widely used cross-entropy loss (which is used for the slide-level classification task), we apply the binary top-1 smooth SVM loss to the outputs of the clustering layers of the network. In all our experiments, $\alpha$ and $\tau$ are both set to $1.0$.
\begin{equation} \label{svmloss}
l(\mathbf{s}, y)=\max \left\{\max _{j \in \mathcal{Y} \backslash\{y\}}\left\{s_{j}+\alpha \right\}-s_{y}, 0\right\}
\end{equation}
\vspace{-5mm}
\begin{equation} \label{smoothsvmloss}
\mathcal{L}_{1, \tau}(\mathbf{s}, y)=\tau \log \left[\sum_{j \in \mathcal{Y}} \exp \left(\frac{1}{\tau}\left(\alpha \mathbbm{1}(j \neq y) + s_{j} - s_{y} \right)\right)\right]
\end{equation}
\\
\textit{Training Details.} During training, slides are randomly sampled and provided to the model using a batch size of 1. The multinomial sampling probability of each slide is inversely proportional to the frequency of its ground truth class (\textit{i.e.} slides from underrepresented classes are more likely to be sampled relative to others) in order to mitigate class imbalance in the training set. The total loss for a given slide $\mathcal{L}_{total}$ is the sum of both the slide-level classification loss $\mathcal{L}_{slide}$ and the instance-level clustering loss $\mathcal{L}_{patch}$ with optional scaling via scalar $c_1 \text{ and } c_2$:
\vspace{-4mm}
\begin{equation}
\mathcal{L}_{total} = c_1\mathcal{L}_{slide} + c_2\mathcal{L}_{patch}
\end{equation}
To compute $\mathcal{L}_{slide}$, $\mathbf{s}_{slide}$ is compared against the ground truth slide-level label using the standard cross-entropy loss and to compute $\mathcal{L}_{patch}$ the instance-level clustering prediction scores $\mathbf{p}_{k}$ for each sampled patch are compared against their corresponding pseudo cluster labels using the binary smooth SVM loss (recall for non-subtyping problems there are a total of $2B$ patches sampled from the in-the-class branch while for subtyping problems there are $2B$ patches sampled from the in-the-class branch and $B$ patches sampled via each of $n-1$ out-of-the-class attention branches). We used $B = 8$ and weights $c_1=0.7$, $c_2=0.3$ for all experiments. The model parameters are updated via the Adam optimizer with an L2 weight decay of 1e-5 and a learning rate of 2e-4. \\
\noindent\textit{Model Selection.}
All models are trained for at least 50 epochs and up to a maximum of 200 epochs if the early stopping criterion is not met. Namely, the validation loss is monitored each epoch and when it has not decreased from the previous low for over 20 consecutive epochs, early stopping is used. The saved model, which has the lowest validation loss, is then tested on the test set. \\
\noindent\textbf{Computational Hardware and Software} \\
We used multiple hard drives to store the raw files of digitized whole slides. Segmentation and patching of WSIs are performed on Intel Xeon CPUs (Central Processing Units) and feature extraction using a pretrained neural network model is accelerated through data batch parallelization across multiple NVIDIA P100 GPUs (Graphics Processing Units) on Google Cloud Compute instances or 2080 Ti GPUs on local workstations. All weakly-supervised deep learning models are trained with a total of 10 local, consumer workstation-grade NVIDIA 2080 Ti GPUs by streaming extracted features from fast local SSD (Solid State Drive) storage. Exactly 2 GPUs are used for training in each experiment. Our whole slide processing pipeline is implemented in Python and takes advantage of imaging processing libraries such as openslide (version 3.4.1), opencv (version 4.1.1) and pillow (version 6.2.1). For loading data and training deep learning models using CLAM, we used the Pytorch (version 1.3) deep learning library. All plots were generated using matplotlib (version 3.1.1) and seaborn (version 0.8.1). Area under the curve of the receiver operating characteristic curve (AUC ROC) was estimated using the Mann-Whitney U-statistic, for which the algorithmic implementation is provided in the scikit-learn scientific computing library (version 0.22.1). The 95\% confidence intervals of the true AUC (included in \textbf{Supplementary Table 4-10, 13-15} was estimated using DeLong's method.
\\
\textbf{WSI Datasets} \\
All in-house slides are digitized at BWH using a Hamamatsu S210 scanner at the 40$\times$ magnification. A summary of all datasets used are included in \textbf{Supplementary Table 11}. \\
\textit{External RCC WSI Dataset.} 
Our external renal cell cancer dataset consists of a total of 884 diagnostic WSIs from the TCGA Kidney repository under the Kidney Chromophobe (TGCA-KICH), Kidney Renal Clear Cell Carcinoma (TCGA-KIRC) and Kidney Renal Papillary Cell Carcinoma (TCGA-KIRP) projects. There are 111 Chromophobe slides from 99 cases, 489 Clear Cell slides from 483 cases, and 284 Papillary Cell slides from 264 cases. The mean number of patches extracted per slide at 20$\times$ magnification is 13907. \\
\textit{In-house BWH RCC WSI Dataset.}
Our internal renal cell cancer dataset consists of a total of 135 WSIs from 133 cases of which 43 slides are Chromophobe, 46 are Clear Cell and 46 are Papillary Cell. The mean number of patches extracted per slide at 20$\times$ magnification is 20394. Our renal cell cancer biopsy dataset consists of a total of 92 WSIs from 79 cases, of which 13 slides are Chromophobe, 53 slides are Clear Cell, 26 are Papillary Cell. The mean number of patches extracted per slide at 20$\times$ magnification is 1709. Our kidney cellphone dataset comprises of 4-6 FOVs per slide for each of the 135 slides. The mean number of patches extracted for each set of FOVs is 419. All slides were collected and processed at the Brigham and Women's Hospital between year 2010 and 2019.\\
\textit{External Lung WSI Dataset.} 
Our external non-small cell lung cancer dataset consists of 993 diagnostic WSIs from the TCGA Lung repository under the Lung Squamous Cell Carcinoma (TGCA-LUSC) and Lung Adenocarcinoma (TCGA-LUAD) projects. There are 507 LUAD slides from 444 cases and 486 LUSC slides from 452 cases. Additionally, we collected a total of 1526 WSIs from the TCIA CPTAC Pathology Portal that have lung as the topological site. From these WSIs, 668 slides from 223 cases are labeled as LUAD and 306 slides from 108 cases are labeled as LUSC. The remaining 552 slides are labeled as normal tissue and excluded. Accordingly, in total, our external lung carcinoma dataset contains 1967 WSIs (1175 LUAD slides from 667 cases and 792 LUSC cases from 560 patients). The mean number of patches extracted per slide at 20$\times$ magnification is 9958. \\
\textit{In-house BWH NSCLC WSI Dataset.}
Our internal non-small cell lung cancer dataset consists of a total of 131 resection slides (63 LUAD slides and 68 LUSC slides) and 110 biopsy slides (55 LUAD slides and 55 LUSC slides). Each slide comes from a unique case. The mean number of patches extracted per biopsy slide and per resection slide at 20$\times$ magnification is 820 and 24714 respectively. All slides were collected and processed at BWH between year 2016 and 2019. Our lung cellphone dataset comprises of 4-8 FOVs per slide for each of the 131 resection slides. The mean number of patches extracted for each set of FOVs is 406. \\
\textit{External Lymph Node WSI Dataset.}
Camelyon16 and Camelyon17\cite{camelyon} are two of the largest publicly available, annotated breast cancer lymph node metastasis detection datasets. Camelyon16 consists of 270 annotated whole slides for training and another 129 slides as a held-out, official test set collected at the Radboud University Medical Center and the University Medical Center Utrecht in the Netherlands. On the other hand, Camelyon17 consists of a total of 1000 slides from 5 different medical centers in the Netherlands. Because slide-level labels for the 500 slides in the official test set of Camelyon17 were not yet publicly available, we used just the training portion of Camelyon17, which consists of 500 slides (with corresponding slide-level diagnosis) for 100 cases. We combined Camelyon16 and Camelyon17 into a single dataset with a total of 899 slides (591 negative and 308 positive) from 370 cases. The mean number of patches extracted per slide at 40$\times$ magnification is 41802. \\
\textit{In-house BWH Lymph Node Met. (Breast Cancer) WSI Dataset.}
Our internal breast cancer lymph node metastasis dataset consists of a total of 133 WSIs from 131 cases (66 negative slides and 67 positive slides). The mean number of patches extracted per slide at 40$\times$ magnification is 51426. These slides were collected at BWH between year 2017 and 2019. \\
\noindent\textbf{WSI Processing} \\
\textit{Segmentation.} For each digitized slide, our pipeline begins with automated segmentation of the tissue regions. The WSI is read into memory at a downsampled resolution (\textit{e.g.} $32\times$), converted from RGB to the HSV color space. A binary mask for the tissue regions (foreground) is computed based on thresholding the saturation channel of the image after median blurring to smooth the edges, and is followed by additional morphological closing to fill small gaps and holes. The approximate contours of the detected foreground objects are then filtered based on an area threshold and stored for downstream processing while the segmentation mask for each slide is made available for optional visual inspection. A human-readable text-file is also automatically generated, which includes the list of files processed along with editable fields containing the set of key segmentation parameters used. While the default set of parameters are generally sufficient for reliable tissue segmentation, they can also easily be manually edited for any individual slide should the user find its segmentation results unsatisfactory. \\
\textit{Patching.}
After segmentation, for each slide, our algorithm exhaustively crops 256 $\times$ 256 patches from within the segmented foreground contours at the user specified magnification and stores stacks of image patches along with their coordinates and the slide metadata using the hdf5 hierarchical data format. Depending on the size of each WSI and the specified magnification, the number of patches extracted from each slide can range from hundreds (biopsy slide patched at 20$\times$ magnification) to hundreds of thousands (large resection slide patched at 40$\times$ magnification). 
\\
\textit{Feature Extraction.}
Following patching, for each slide, we use a deep convolutional neural network to compute a low-dimensional feature representation for each image patch. Namely, we take a ResNet50 model pretrained on ImageNet and use adaptive mean-spatial pooling after the 3rd residual block of the network to convert each 256 $\times$ 256 patch into a 1024-dimensional feature vector using a batch size of 128 per GPU. The benefits of using extracted features as inputs to deep learning models for supervised learning include drastically faster training time and lower computational cost. This enables us to train a deep learning model on thousands of WSIs in a matter of couple hours once the features have been extracted. Compared to using raw pixels, using low-dimensional features also makes it feasible to fit all patches in a slide (up to 150,000 or more) into GPU memory simultaneously, thus avoiding the need for sampling patches and using noisy labels. \\
\newpage
\noindent\textbf{Visualization} \\
\textit{Visualizing Slide-level Feature Space.} For each public WSI dataset, a model trained on one of the ten training sets created for cross-validation was used to compute a 512-dimensional slide-level feature representation for every slide in its corresponding validation and test set based on the model's final slide-level prediction. The resulting set of slide-level feature vectors are reduced to 2-dimensional space through transformation via PCA (Principle Component Analysis) with 50 components, visualizing the first and second principle component, and shading each point by its ground truth slide-level label. We then repeated this procedure for the models trained on 10\%, 25\%, 50\% and 75\% of the same training set. We also performed the same analysis on the slides in each independent test cohort using the best performing model for each training set size. \\
\textit{Interpreting Model Prediction via Attention Heatmap.} In order to interpret the relative importance of different regions in a slide to the model's final slide-level prediction, we computed and saved the unnormalized attention scores (before they are converted to probability distribution by applying the softmax function) for all the patches extracted from the slide, using the attention branch that corresponds to the model's predicted class. These attention scores were converted to percentile scores and scaled to between 0 and 1.0 (with 1.0 being most attended and 0.0 being least attended). The normalized scores are converted to RGB colors using a diverging colormap and displayed on top of their respective spatial locations in the slide to visually identify and interpret regions of high attention displayed in red (positive evidence, high contribution to model's prediction relatively to other patches) and low attention displayed in blue (negative evidence, low contribution to model's prediction relatively to other patches). To create more fine-grained heatmaps, we tile the slides or smaller ROIs (\textit{e.g.} 8000 $\times $ 8000) into 256 $\times $ 256 patches using an overlap and calculate the raw attention score for each patch. We then followed a similar procedure and used the same colormap as above to convert the raw score of each patch in the ROI to RGB colors. To ensure that the normalized attention scores computed for patches produced with an overlap are directly comparable to those for the set of non-overlapping patches used by the model for prediction, we referred to the set of unnormalized attention scores over the entire slide (without overlap) when calculating each patch's percentile score. The ROI heatmaps are overlaid over the original WSI with a transparency value of 0.5. in order to simultaneously visualize the underlying morphological structures in the original H\&E slide. ROI and biopsy heatmaps are produced with an overlap of 95\%. To produce fine-grained heatmaps for cellphone images, a 95\% overlap was used and attention scores were normalized over each image. \\
\textit{Visualizing Patch-level Feature Space.} For each slide in the independent test cohort, we uniformly random sampled 2\% of its tissue patches and recorded their clustering probability predictions made by each of the $n$ clustering branches in addition to their 512-dimensional feature representations after the first fully-connected layer. For subtyping problems, patches for which all clustering branches predicted a positive probability of less than 0.5 (in other words, the clustering branch of every class considers them as negative evidence for its respective class) were labeled as class-agnostic, while the remaining patches were labeled with the class for which its positive probability is the highest. For metastasis detection in axillary lymph nodes, the clustering branch corresponding to the positive class is instead used to label patches as positive (positive probability greater than 0.5) and class-agnostic (positive probability less than 0.5). Using the same technique above for visualizing the slide-level feature space, we reduced each patch-level feature vector to two dimensions using PCA. \\
\noindent\textbf{Comparative Analysis} \\
\textit{Multiple Instance Learning.} The most well-known MIL decision rule involves a diagnostic model making a prediction for every patch in a whole slide and the patch with the highest predicted probability for the positive class is selected to both inform the final diagnostic decision for the entire slide as well as gradient signals during training. In addition to using MIL, which simply takes the highest probability patch, Campanella \textit{et al}.\cite{campanella2019clinical} also introduced an RNN-based aggregation that sequentially passes the top $S$ patches ranked based on their predicted probability for the positive class through an RNN to obtain the final slide-level prediction. However, out of the three datasets for which they compared RNN-based aggregation to MIL-alone, they only noted a statistically significant improvement in test AUC of 0.5\% on the prostate dataset, whereas no statistically significant improvement was found for the classification of skin cancer basal cell carcinoma (BCC) and lymph node metastasis detection. In light of these findings, we implemented the classic MIL formulation as our baseline for comparison.
\\
\textit{Binary Implementation.}
For each slide, during training, feature embeddings of all patches in the slide are read into memory at once, which corresponds to an input into the MIL network of shape $K \times 1024$. $K$ is the number of patches, which varies from slide to slide (known as the bag size) and each patch is described by a fixed 1024-dimensional vector representation $\mathbf{z}_{k}$, produced previously in the feature extraction step using a pretrained ResNet50 model. The MIL network has one fully-connected layer with 512 hidden units and is followed by the classification layer. If we denote the weights and bias of each layer as $\mathbf{W}_{1} \in \mathbbm{R}^{512 \times 1024}$, $\mathbf{b}_{1} \in \mathbbm{R}^{512}$ and $\mathbf{W}_{2} \in \mathbbm{R}^{2 \times 512}$, $\mathbf{b}_{2} \in \mathbbm{R}^{2}$ respectively, the unnormalized prediction score $\mathbf{s}_{k}, 1 \leq k \leq K$ for each patch can therefore be defined as:
\vspace{-4mm}
\begin{equation}
\mathbf{s}_{k}= \mathbf{W}_{2}(\text{ReLU}(\mathbf{W}_{1}\mathbf{z}_{k}^{\top} + \mathbf{b}_{1})) + \mathbf{b}_{2}
\end{equation}
According to the max pooling aggregation rule, the patch whose predicted score for the positive class is the highest is then selected to represent the final slide-level prediction.
\textit{Multi-class Implementation.}
As previously mentioned, the MIL algorithm was designed specifically for binary classification. In order to compare the performance of CLAM against MIL in the multi-class setting, we implemented a multi-class variant of MIL which we call mMIL. mMIL has a fully-connected layer of the same dimension as our binary MIL network but we adjust the binary classification layer to be instead $\mathbf{W}_{2} \in \mathbbm{R}^{n \times 512} $ in order to predict the $n$-class probability distribution of every patch in the slide. Similar to performing max pooling in the binary case, based on the raw scores, we select the patch with the highest single class score as the slide-level prediction. \\
\textit{Training Details.}
During training, scores of the patch selected via max pooling is passed to the cross-entropy loss function and the model parameters are optimized via stochastic gradient descent using the Adam optimizer with a learning rate of 2e-4 and weight decay of 1e-5. We use the same mini-batch sampling strategy as well as the same early stopping and model selection criteria for MIL/mMIL as for CLAM. For inference, the predicted probability distribution over each class is computed by normalizing the raw predicted scores of the max-pooled patch using a softmax function. 
\vspace{-8mm}
\section*{Data Availability}
\vspace{-8mm}
The TCGA diagnostic whole slide data (lung, kidney) and corresponding labels are available from NIH genomic data commons and CPTAC whole slide data (lung) is available from the NIH cancer imaging archive. Metastatic lymph node data are publicly available from the Camelyon 16 and 17 websites. We included links to all data sources in \textbf{Supplementary Table 12}. Reasonable requests for in house BWH data may be addressed to the corresponding author.
\vspace{-8mm}

\section*{Code Availability}
\vspace{-8mm}
All code was implemented in Python using PyTorch as the primary deep learning package. Code, trained models and scripts to reproduce the experiments of this paper are available at \url{https://github.com/mahmoodlab/CLAM}
All source code is provided under the GNU GPLv3 free software license.
\vspace{-8mm}

\section*{Author Contributions}
\vspace{-8mm}
M.Y.L. and F.M. conceived the study and designed the experiments. M.Y.L. performed the experimental analysis. D.W. T.C. curated the in-house datasets, and collected cellphone microscopy data. M.Y.L R.J.C and M.B. developed and tested the CLAM Python package. M.Y.L. F.M. prepared the manuscript. F.M. supervised the research.
\vspace{-8mm}
\section*{Acknowledgements}
\vspace{-8mm}
The authors would like to thank Alexander Bruce for scanning internal cohorts of patient histology slides at BWH; Jingwen Wang, Katerina Bronstein, Lia Cirelli and Sharifa Sahai for querying the BWH slide database and retrieving archival slides; Martina Bragg and Terri Mellen for logistical support; and Zahra Noor for developing the interactive demo website. This work was supported in part by internal funds from BWH Pathology, Google Cloud Research Grant and Nvidia GPU Grant Program. R.J.C. was additionally supported by the NSF Graduate Fellowship and NIH T32HG002295. 
\vspace{-8mm}

\section*{Competing Interests}
\vspace{-8mm}
The authors declare that they have no competing financial interests.
\vspace{-8mm}

\section*{Ethics Oversight}
\vspace{-8mm}
The study was approved by the Mass General Brigham (MGB) IRB office under protocol 2020P000233.
\end{spacing}
\newpage
\begin{nolinenumbers}
\vspace{-9mm}

\section*{References} 
\vspace{2mm}

\begin{spacing}{0.9}
\bibliographystyle{naturemag}
\bibliography{sample}
\end{spacing}
\end{nolinenumbers}
\begin{figure*}
\includegraphics[width=\textwidth]{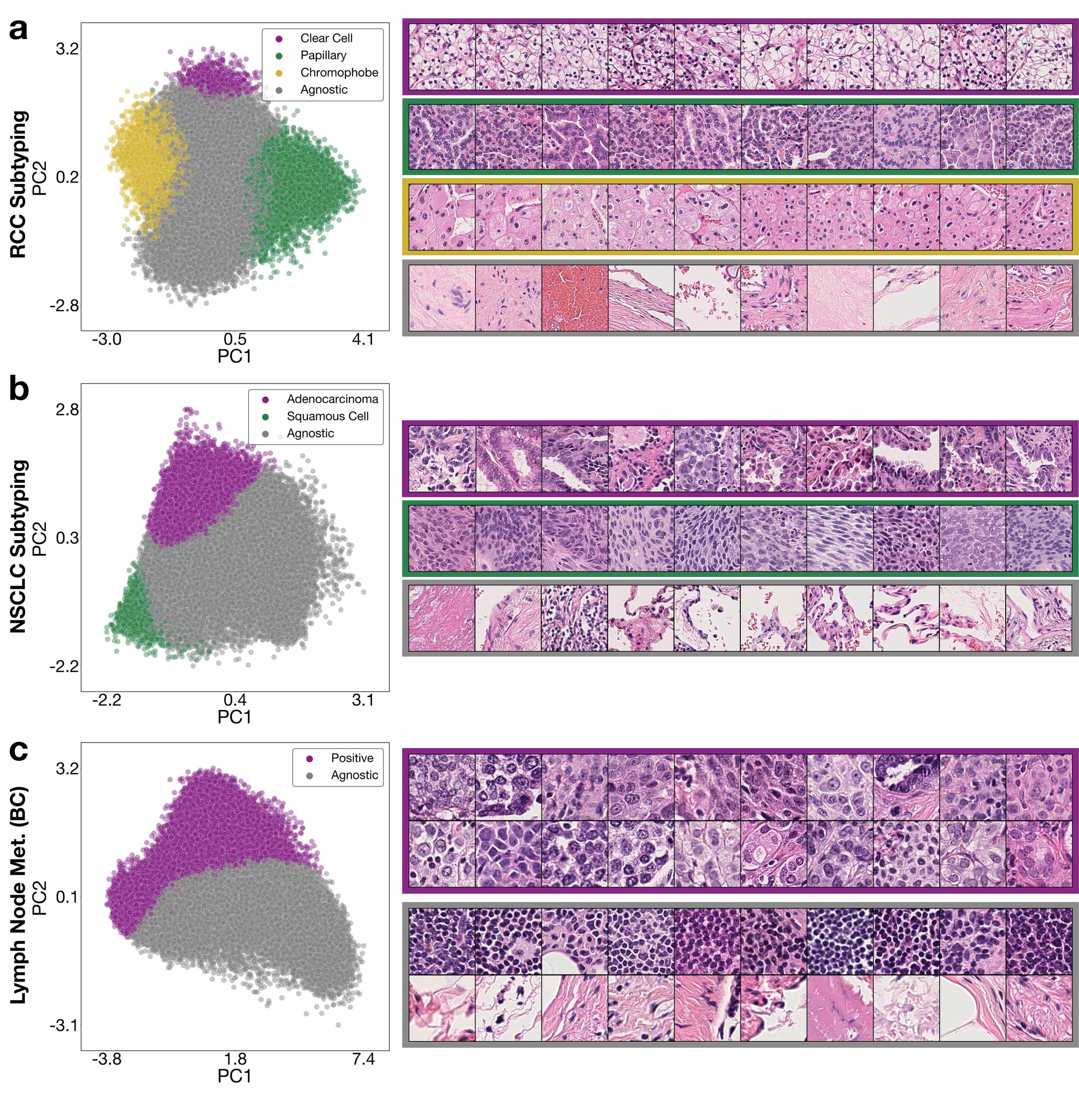}
\caption*{\textbf{Extended Data Figure 1. Visualizing the patch-level feature space.} To visualize the patch-level feature space, for each task, we randomly sampled 2\% of patches from each slide in the independent test cohort and reduced their 512-dimensional feature representation to two dimensions using PCA (left). For subtyping tasks (\textbf{a, b}), each patch is shaded with the class predicted ($p  \geq 0.5$) by the clustering layers of the model. If a patch is predicted as negative ($p < 0.5$) for all classes, it is labeled as 'Agnostic'. We observe that patches predicted as different subtypes are separated into distinct clusters in the feature space, and patches sampled from each cluster generally exhibit morphology characteristic of each subtype. Similarly, for metastasis detection in axillary lymph nodes (\textbf{c}), patches are shaded as positive ($p \geq 0.5$ for the positive class) and agnostic ($p < 0.5$). Tumor cells are clearly picked out by the positive cluster and the agnostic cluster corresponds with immune cells and normal tissue.}
\end{figure*}

\begin{figure*}
\includegraphics[width=\textwidth]{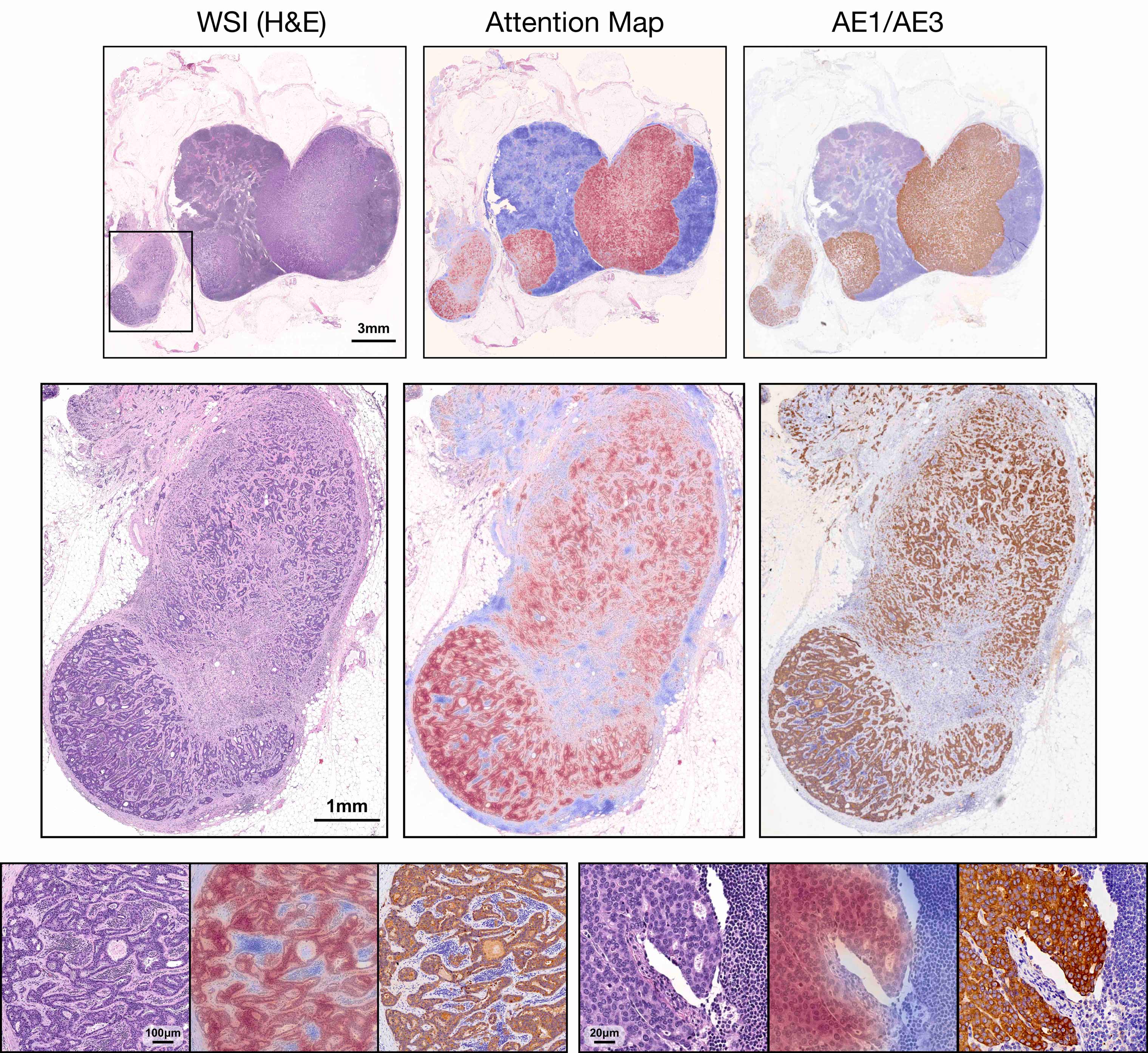}
\caption*{\textbf{Extended Data Figure 2. Validation of attention heatmap for axillary lymph node metastasis using cytokeratin (AE1/AE3) immunohistochemical staining.}  Subsequent slices of paraffin-embedded tissue of several positive cases of axillary lymph node metastasis are collected, cut and stained with H\&E and AE1/AE3 IHC, and digitized at BWH. A CLAM model trained on our public lymph node metastasis training set is tested on the entire tissue region (excluding fat) of the H\&E WSI using overlapping patches and a fine-grained attention heatmap corresponding to the model's prediction is created. We find that in addition to correctly detecting metastasis at the slide-level, CLAM is capable of accurately attending to metastatic regions (red in attention heatmap, gold in corresponding IHC) and even individual tumor cells in the side-by-side comparison of the fine-grained attention heatmap and IHC-stained WSI. This promising finding signals that in some circumstances, it might be possible to apply CLAM (which requires no pixel-level or ROI-level annotation and no special stains for training) to whole-slide-level segmentation tasks (including but not limited to predicting the corresponding IHC) that would otherwise incur either costly labour and human expertise or expensive reagents and core facilities.}
\end{figure*}

\begin{figure*}
\includegraphics[width=\textwidth]{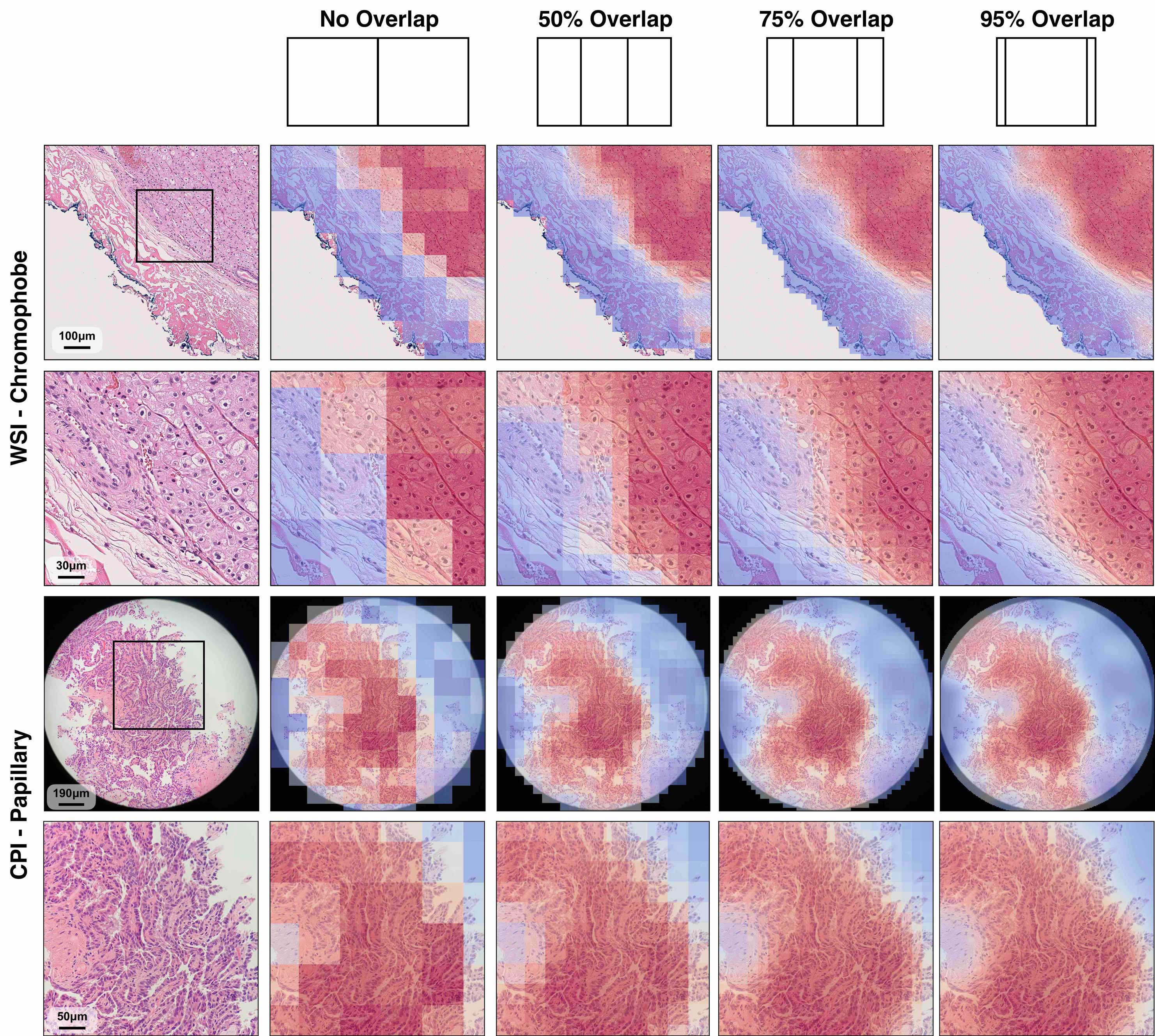}
\caption*{\textbf{Extended Data Figure 3. Attention heatmap visualization using varying overlap.} In our study, CLAM uses 256 $\times$ 256 patches to make predictions. By default, patches cover the entire extent of the tissue regions in each slide with a step size of 256 (no overlap) for fast training and inference. As a result, the attention heatmap appears blocky as there can be large transitions in the attention scores assigned to  neighboring patches. Instead of using interpolation techniques to estimate the attention scores for overlapped locations that are not sampled during patching, we propose to increase the overlap between patches (up to 95\% overlap) for fine-grained heatmap visualization. Attentions scores are first normalized to percentile scores by referring to the raw scores computed for all unoverlapped patch locations (this ensures that the same locations from the overlapped and nonoverlapped heatmaps always have roughly the same normalized scores). After normalization, attention scores are mapped to their corresponding spatial locations in the WSI and visualized (scores for overlapped regions are accumulated and averaged). As demonstrated in both the WSI and CPI example, using an overlap above 50\% significantly reduces the blockiness of the resulting heatmap and using a 95\% overlap renders the heatmap nearly completely smooth to a human observer.}
\end{figure*}

\begin{figure*}
\includegraphics[width=\textwidth]{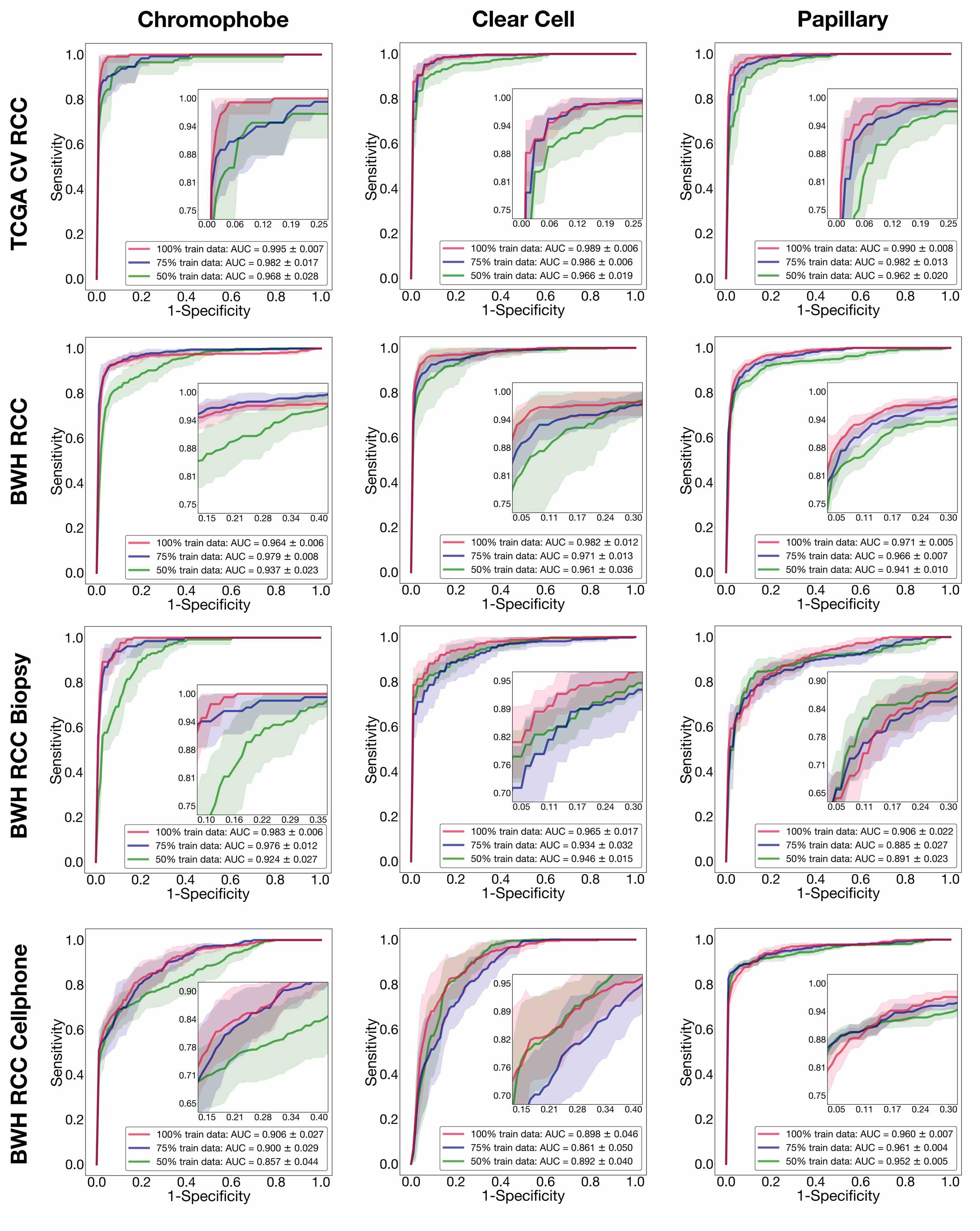}
\vspace{-8mm}
\caption*{\textbf{Extended Data Figure 4. One-vs-rest ROC curve for RCC subtyping.} ROC curve is drawn for each of the three classes (Chromophobe, Clear Cell and Papillary) for each RCC subtyping dataset (TCGA kidney dataset, BWH independent test set, BWH biopsy dataset, and BWH cellphone dataset) by considering the probability predictions and ground truth labels for the 3-class classification problem as one-vs-rest. }
\end{figure*}

\begin{figure*}
\includegraphics[width=\textwidth]{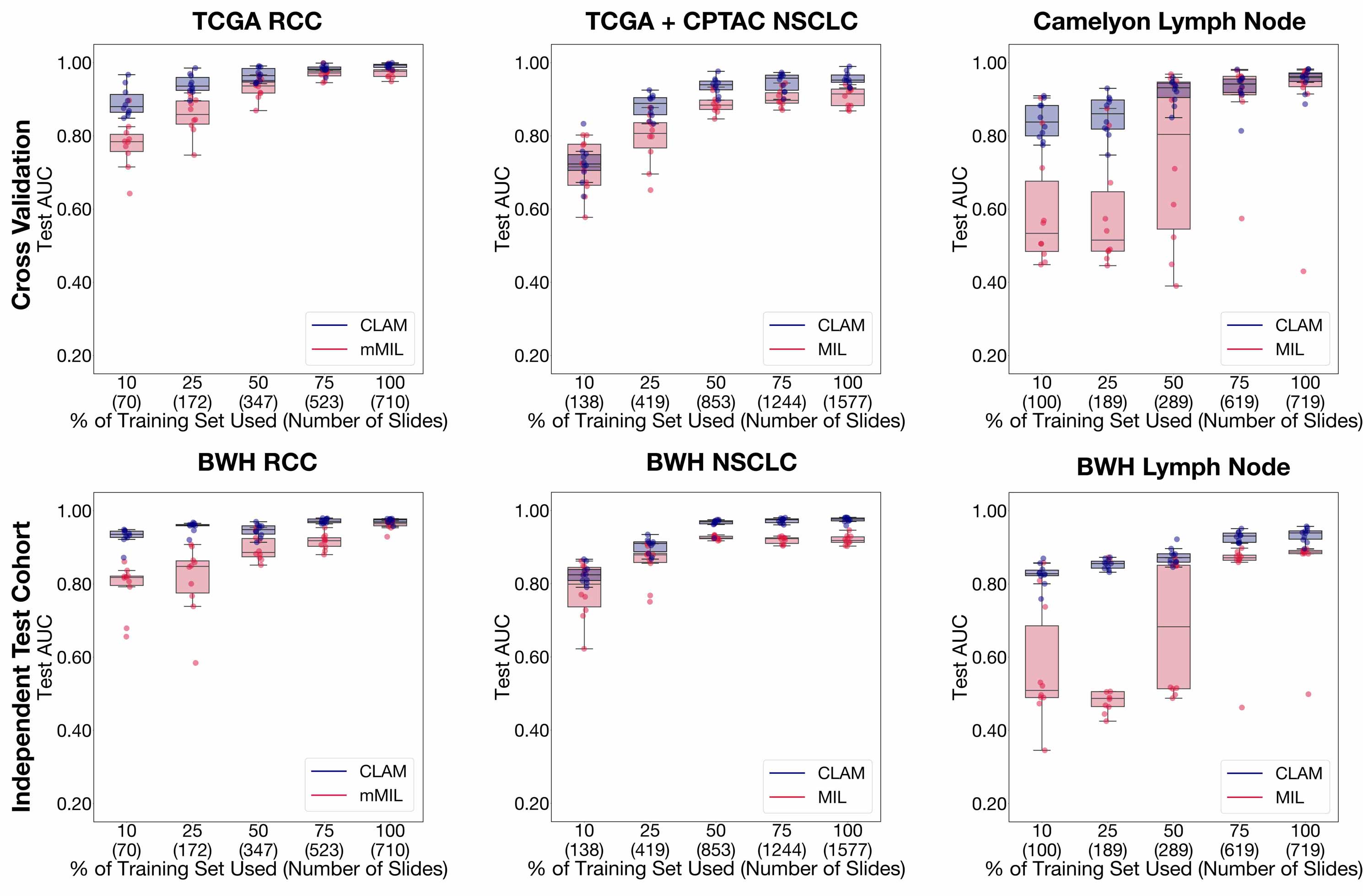}
\vspace{-8mm}
\caption*{\textbf{Extended Data Figure 5. Comparison with MIL/mMIL.} CLAM is compared against MIL in the two-class classification tasks (axillary lymph node metastasis detection and NSCLC subtyping) and against mMIL in the 3-class RCC subtyping task. For all tasks, the exact same dataset partitioning is used in cross-validation and the same training hyperparameters are used for both algorithms (learning rate, weight decay, optimizer, early stopping criteria, \textit{etc.}). In the comparison plots, for each training set size, the test AUC of each trained model is shown as a single dot in a vertical strip plot and the distribution of all 10 trained models (there is one trained model for each algorithm for each of 10 different cross-validation partitions) is summarized by a boxplot. In RCC subtyping, the performance is comparable between mMIL and CLAM in both cross-validation and on the independent test set when using all the available training data but the gap widens towards the smaller training set sizes, demonstrating that CLAM is more data efficient. In NSCLC subtyping CLAM outperforms MIL for most dataset sizes by 3-5\% on average. In lymph node metastasis detection, MIL failed to generalize in a number experiments, scoring an AUC of under or close to 0.5 on the cross-valitation test sets and the independent test set, signaling that MIL is potentially very sensitive to the sampled training data distribution, hyperparamters and/or the stochasticity in the network's initialization, data sampling during training, \textit{etc.} This phenomenon is exacerbated when the training set size is small (\textit{e.g.} 25\% and 10\% of training cases) and around half of the MIL models did not converge to a solution capable of generalization. In cross-validation, when MIL did train successfully and generalize on the validation and test set, its performance tend to be comparable to CLAM. However, on our lymph node independent test set, CLAM outperforms MIL consistently across all training set sizes. All results for mMIL/MIL and CLAM on the independent test sets with associated 95\% confidence intervals are included in \textbf{Supplementary Table 4-10}.}
\end{figure*}

\begin{figure*}
\includegraphics[width=\textwidth]{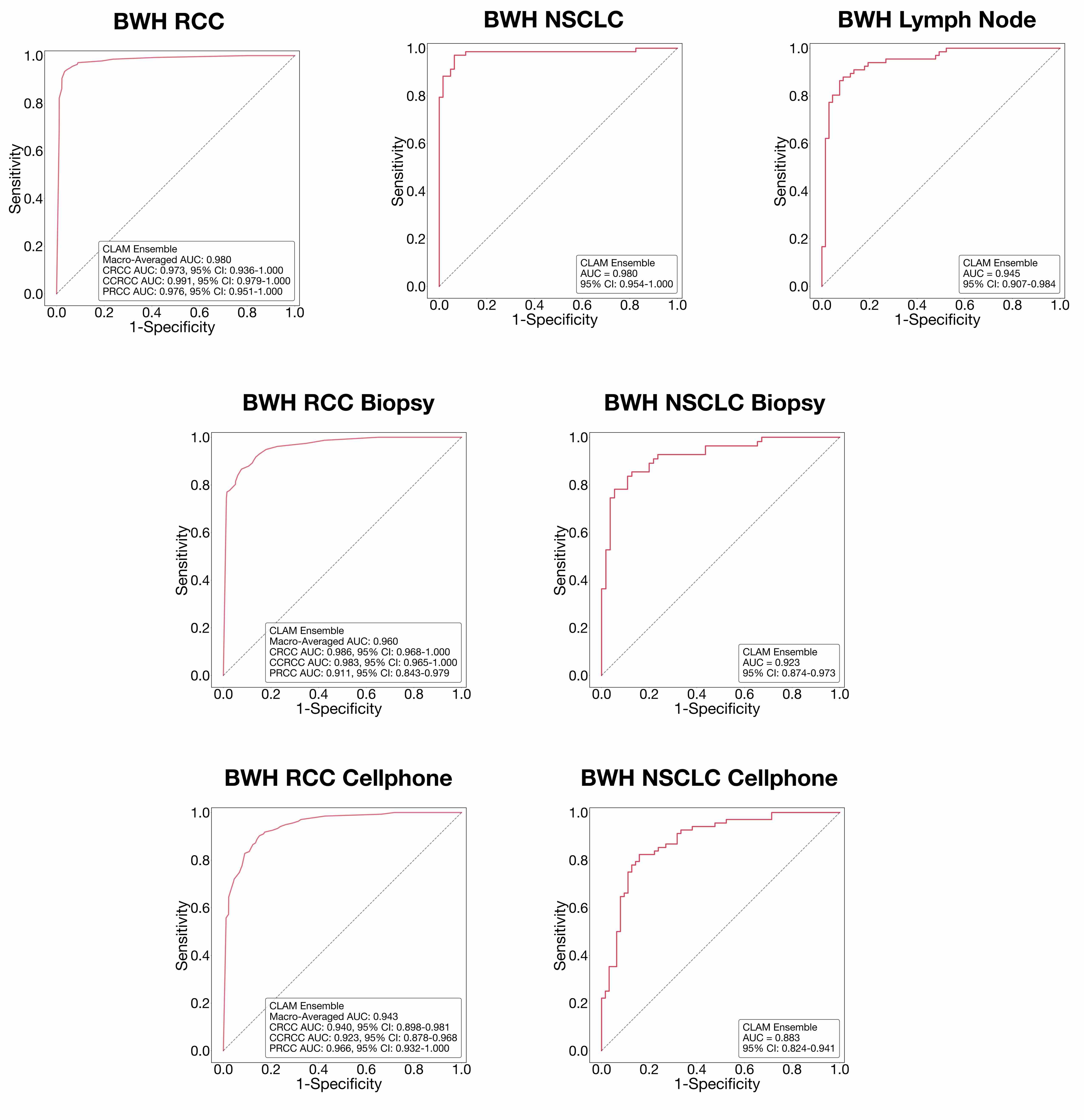}
\vspace{-8mm}
\caption*{\textbf{Extended Data Figure 6. Performance of CLAM ensemble system on independent test sets.} For each task, we took the models trained on the public datasets using 10-fold monte-carolo cross-validation (80\% of cases in each dataset were used to train each model) and computed their ensemble predictions by averaging the softmax-activated probability scores over all 10 readers for each slide in the independent test set. For BWH NSCLC subtyping and axillary lymph node metastasis detection, the test AUC of the ROC curve corresponding the ensemble predictions is computed along with its 95\% confidence interval (CI). For BWH RCC subtyping, the one-vs-rest AUC and its 95\% CI for each subtype is computed in addition to the macro-averaged AUC.}
\end{figure*}

\end{document}